\documentclass[journal = jctcce,manuscript=article,layout=twocolumn]{achemso}

\usepackage{graphicx}
\usepackage{dcolumn}
\usepackage{bm}
\usepackage{color}
\usepackage[utf8]{inputenc}
\usepackage{textalpha}
\DeclareUnicodeCharacter{0301}{ }
\usepackage{xcolor}
\usepackage{subcaption}
\usepackage{comment}
\usepackage{hyperref}
\usepackage{xspace}
\usepackage{listings}
\usepackage{tikz}

\newcommand{\scalebarimg}[5]{
  \centering
  \begin{tikzpicture}
    \draw node[name=micrograph] {\includegraphics[width=#2\linewidth]{#1}}; 
    \draw  (micrograph.north west)  node[anchor=north west,yshift=-#4, xshift=#5,black]{\textbf{\large{#3}}}; 
  \end{tikzpicture}
}

\newcommand{\rct}{RADICAL-Cybertools\xspace}

\title{Extensible and Scalable Adaptive Sampling on Supercomputers}

\author{Eugen Hruska}
\affiliation{Center for Theoretical Biological Physics, Rice University, Houston, TX, USA}
\alsoaffiliation{Department of Physics \& Astronomy, Rice University, Houston, TX, USA}

\author{Vivekanandan Balasubramanian}
\affiliation{Department of Electrical and Computer Engineering, Rutgers University, Piscataway, NJ, USA}

\author{Hyungro Lee}
\affiliation{Department of Electrical and Computer Engineering, Rutgers University, Piscataway, NJ, USA}

\author{Shantenu Jha}
\affiliation{Department of Electrical and Computer Engineering, Rutgers University, Piscataway, NJ, USA}

\author{Cecilia Clementi}
\affiliation{Department of Physics, Freie Universit\"at, Berlin, Germany}
\alsoaffiliation{Center for Theoretical Biological Physics, Rice University, Houston, TX, USA}
\alsoaffiliation{Department of Chemistry, Rice University, Houston, TX, USA}
\alsoaffiliation{Department of Physics \& Astronomy, Rice University, Houston, TX, USA}
\email{cecilia.clementi@fu-berlin.de}

\date{\today}
\begin{document}

\begin{abstract}
The accurate sampling of protein dynamics is an ongoing challenge despite the
utilization of High-Performance Computers (HPC) systems. Utilizing only
``brute force'' MD simulations requires an unacceptably long time to solution.
Adaptive sampling methods allow a more effective sampling of protein dynamics
than standard MD simulations. Depending on the restarting strategy the speed
up can be more than one order of magnitude. One challenge limiting the
utilization of adaptive sampling by domain experts is the relatively high
complexity of efficiently running adaptive sampling on HPC systems. We
discuss how the ExTASY framework can set up new adaptive sampling strategies,
and reliably execute resulting workflows at scale on HPC platforms. Here the
folding dynamics of four proteins are predicted with no a priori
information.
\end{abstract}

\maketitle

\definecolor{background}{rgb}{0.94,0.95,0.96}
\lstset{ 
 basicstyle=\ttfamily\scriptsize, 
 breakatwhitespace=false, 
 breaklines=true, 
 backgroundcolor=\color{background},
 captionpos=b, 
 commentstyle=\color{gray}, 
 escapeinside={\%*}{*)}, 
 extendedchars=true, 
 frame=single, 
 keepspaces=true, 
 keywordstyle=\color{blue}, 
 language=Python, 
 numbers=none, 
 numbersep=5pt, 
 numberstyle=\tiny\color{gray}, 
 showspaces=false, 
 showstringspaces=false, 
 showtabs=false, 
 stepnumber=2, 
 stringstyle=\color{green}, 
 tabsize=2, 
 moredelim=**[is][\color{red}]{@}{@},
 xleftmargin=1em,
 framexleftmargin=-1.5em,
}

\section{\label{sec:intro}Introduction}

Molecular dynamics (MD) simulations with all-atom force-fields allow simulating
protein folding and protein kinetics with good accuracy. Reaching biologically
relevant processes, such as protein folding or drug binding, is limited mainly
by the required large computational resources and long simulation times. The
long simulation times can be reduced either by simulating parallel trajectories
with massively-distributed computing \cite{DistComp-Shirts2000,
DistComp-Buch2010} or with special-purpose hardware \cite{shaw2014anton}.
Further reduction of required computational resources or simulation times would
allow a more broad application of MD simulations.

One method of reducing both the computational resources and the simulation
times is \emph{adaptive sampling} \cite{singhal2005error, bowman2010enhanced,
weber2011characterization, Fabritiis-2014, preto2014fast, doerr2016htmd,
AdaptivePELE-Lecina2017, 
weexplore, Adstrategies2018, FUNN, FAST, shkurti2019jctc, harada2015jctc, harada2017jctc,
EvolutionCoupling-Shamsi2017, FAST-Bowman-2015, 
Strategies-erros-reduce, plattner2017complete}
Adaptive sampling is an iterative process, where MD simulations from previous
iterations are analyzed, and, based on the analysis, a new iteration of relatively
short MD trajectories is initiated. The starting conformations for the
MD trajectories are determined in such a way to efficiently
reach a goal such as crossing rare transitions barriers, folding a protein, or
recovering the dynamics of a macromolecule. The exact strategy where to restart
new MD simulations determines the success of the adaptive sampling approach,
and several different methods have been proposed and investigated\cite{Fabritiis-2014,
AdaptivePELE-Lecina2017, preto2014fast, doerr2016htmd,
weexplore, Adstrategies2018, FUNN, FAST, shkurti2019jctc}. 
Adaptive sampling
requires to use multiple parallel simulations and is therefore suitable for
High-Performance Computers (HPC).

Determining the efficiency, accuracy, and reliability of a particular adaptive
sampling strategy is challenging for several reasons. Different proteins can
behave differently for different adaptive sampling strategies but limited
computational resources don't allow to adaptively sample a statistically
significant number of proteins with different strategies for comparison.
Accurate results are known only for a limited number of proteins. Despite
these challenges some performance analyses of adaptive sampling strategies
have been reported
\cite{preto2014fast,weber2011characterization,bowman2010enhanced,Fabritiis-2014,
Adstrategies2018}. The results show that some adaptive sampling strategies are
both reliable, accurate and reach speed ups of one or more orders of magnitude
compared to plain MD. For larger, more complex proteins a higher speed up is expected \cite{Adstrategies2018}.

An important challenge in adaptive sampling simulations is the complexity of
performing the required computational tasks efficiently on HPC platforms with
heterogeneous software and hardware environments. This complexity can detract
from the core objective of investigating the behavior of a particular protein
or the efficiency of new adaptive sampling strategies. Some of the existing
frameworks which currently strive to reduce this entry-barrier to
adaptive sampling, such as HTMD\cite{doerr2016htmd}, SSAGES \cite{SSAGES}, DeepDriveMD \cite{leeDeepDriveMDDeepLearningDriven2019}
are either bound to specific software packages, algorithms, computing
platforms or are not open source. 
Additional sampling frameworks involving neural
networks \cite{jung2019acp, ribeiro2018tjocp, bonati2019pnasu} have reported
effective sampling of toy systems and small peptides.
Here we discuss the ExTASY \cite{Extasy2016} framework: we show that it works for
proteins significantly larger than small peptides, scales to
1000s of GPUs and is not limited to specific software packages. ExTASY supports
multiple adaptive sampling
algorithms, different dimension reduction methods and is extensible to new
algorithms and methods while being open source and agnostic of the HPC
platform. In addition to a demonstration of the
scientific results that can be achieved by using ExTASY, in this manuscript, we
investigate the advantages of scalability, reliability or reproducibility
arising from the ExTASY framework for adaptive sampling.

\section{\label{sec:methods}Methods}

Many different implementations of adaptive sampling exist but they all 
have in common
that MD trajectories are periodically analyzed during the simulation, and restart points for the next
batch of trajectories are determined from the analysis of the sampled configurational space.
The different implementations mainly differ in the analysis step, and
they can be based on Markov State Models (MSMs) \cite{prinz2011markov,
MSM-Pande-2018,bookmsm,masterequationsMSM,SCHUTTE1999146}, Diffusion Maps
\cite{Coifman7426, rohrdanz2011determination,Zheng2011, Boninsegna2015},
likelihood-based approaches \cite{peters2006obtaining}, cut-based free energy
profiles \cite{krivov2008diffusive}, or neural networks
\cite{Mardt2018,wehmeyer2018time, ribeiro2018reweighted}. 
In this manuscript, we exemplify the
usage of the ExTASY framework with Markov State Models with different restarting
strategies, as described in Section~\ref{sec:restart-strategies}.

\subsection{\label{sec:adaptive-sampling} Adaptive Sampling}

\begin{figure}[h]
  \centering
  \includegraphics[width=0.9\linewidth]{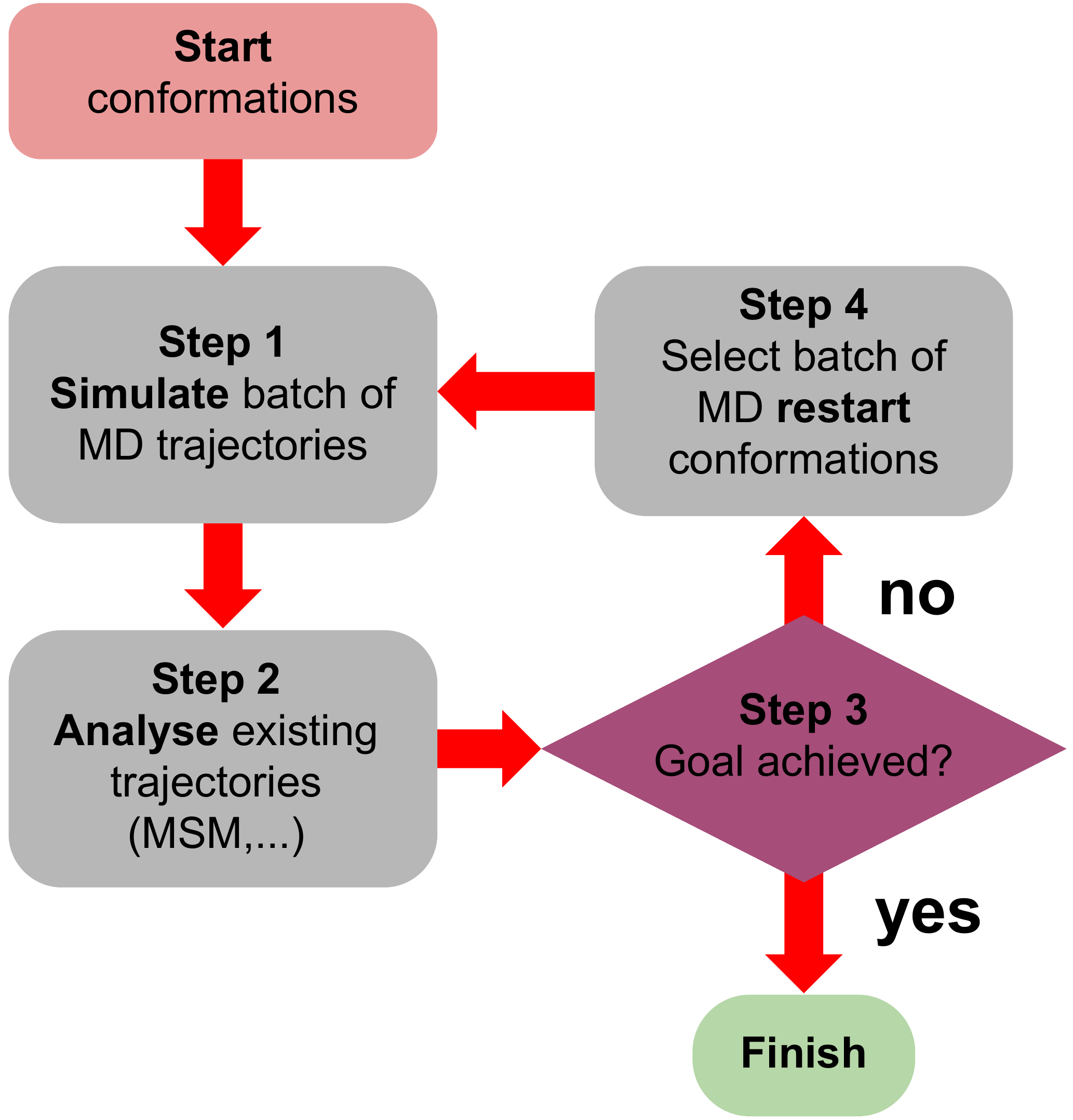}
  \caption{The flow chart shows the basic structure of adaptive sampling. The
  number of starting conformations is variable. The software and hardware
  generating the MD simulations are variable. Different Analysis methods in Step 2
  are possible, commonly TICA \cite{TICA1-perez2013, TICA2-schwantes2013} and
  MSM \cite{prinz2011markov} are used, but alternative methods such as Vampnet \cite{Mardt2018}
  are possible. In Step 3 the goal can be also variable, from finding the whole protein
  dynamics to exploring smaller-scale changes. Step 4 allows different adaptive
  sampling strategies such as the FAST method \cite{FAST} or the strategies
  discussed in this work.}
  \label{fig:schema}
\end{figure}

In each iteration of Markov State Models-based adaptive sampling, all previous
MD simulations are analyzed.
Figure~\ref{fig:schema} is a graphical representation of the process. In the
first iteration of the adaptive sampling, the MD simulations are generated from
a starting state for the system under study, as shown in Step 1.

In Step 2 all previous trajectories are analyzed, first by performing a
dimension reduction. One possible dimension reduction approach is the
Time-lagged Independent Component Analysis (TICA) \cite{TICA1-perez2013,
TICA2-schwantes2013}, that converts the raw trajectories in low-dimensional
trajectories. The Koopman method \cite{koopmanold, koopman2,koopman3,koopman4,
wu2017variational, Nueske2017} can be used to reduce the non-equilibrium effects
emerging from collecting many short MD trajectories and the
resulting low-dimensional trajectories can be scaled into a kinetic map
\cite{Noe2015,noe2016commute}, which provides a measure of the kinetic distance
between different configurations. Another possible dimension reduction approach
implemented (as the default) in ExTASY is based on the state-free
reversible VAMPnets (SRV) \cite{Mardt2018,chen2019jcp}. 
The SRV are essentially a non-linear extension of the TICA method, as they originates from
the same variational approach to conformational dynamics \cite{No2013,Nuske2014}. In SRV, the
trajectories are projected into a low dimensional space by means of a neural network, that allows to model 
non-linearities. 
One practical advantage of the SRV is that they can be accurate even when using
lag times much shorter than in TICA. For adaptive sampling, the shorter lag
time for analysis allows increasing the frequency of restarting
trajectories which potentially improves the efficiency of the sampling. In this case, the
length of MD trajectories in each iteration is reduced and the number of
iterations is increased. 

The dimension reduced trajectories are then clustered with k-means into approximately 200 microstates,
(the exact values for each protein are provided in the Supplementary Information). 
A  maximum-likelihood estimation with a detailed balance
constraint \cite{prinz2011markov} allows obtaining an MSM transition matrix 
between every pair of microstates. The analysis was
performed using the PyEMMA Python package \cite{scherer2015pyemma}, which
allows fast adjustments, and the Hierarchical dynamics encoder package
\cite{chen2019jcp}. The exact parameters for the MSM
construction for each protein are listed in the
Supplementary Information. All these steps can be modified or replaced easily in the
ExTASY workflow. 

\begin{figure}[h]
  \centering
  \includegraphics[width=0.9\linewidth]{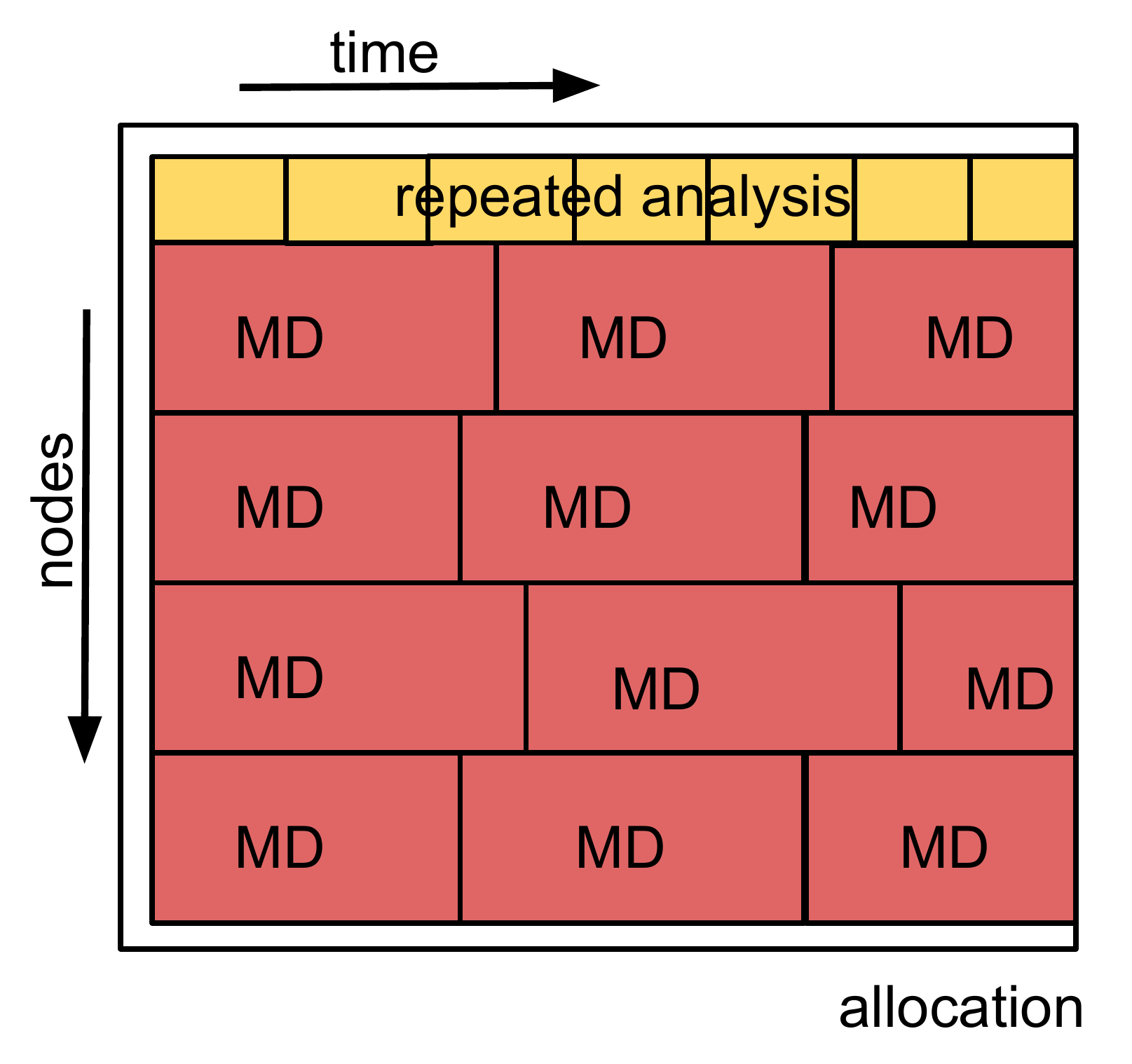}
  \caption{Asynchronous, concurrent execution of the individual molecular
  dynamics and analysis tasks. The Step 4 in Figure~\ref{fig:schema} uses the
  restarting conformations from the latest available analysis results.}
  \label{fig:asynch}
\end{figure}

The overall adaptive sampling process described in Figure~\ref{fig:schema} can be summarized as follows:
\begin{itemize}
\item Start: Start with a start conformation. In the cases presented here, we
start from one unfolded configuration as specified in Section~\ref{sec:MD}.
\item Step 1: Generate a batch of molecular dynamics trajectories from the
selected conformations, the parallelization is defined by the available
computational resources. This step is described in Section~\ref{sec:MD}.
\item Step 2: Analyze the all available data as described in
Section~\ref{sec:adaptive-sampling}, using the probabilities from the MSM
transition matrix.
\item Step 3: Decide if the goal of adaptive sampling is achieved. In this
work, the goal is finding the folded state and obtaining a converged equilibrium
dynamics for the protein. If the goal is not achieved, proceed to Step 4;
otherwise, stop the iterative process.  
\item Step 4: Select the batch of protein conformations for Step 1 in the next
iteration as described in Section~\ref{sec:restart-strategies}.
\end{itemize}

After Step 2 the adaptive sampling continues with Step 4 if the goal is not
achieved. This goal could be folding the protein, achieving a pre-determined
accuracy of the protein dynamics, but could also be manually set to finish
after several iterations. If the goal is not achieved in Step 3, in Step 4 
a batch of protein conformations for the next iteration is selected as
described in Section~\ref{sec:restart-strategies}. If the goal in Step 3 is
achieved, the iterative adaptive sampling finishes and the trajectories can be
further analyzed. 

ExTASY allows to execute the iterative process in Figure~\ref{fig:schema}
in both synchronous and asynchronous manner. In the synchronous case, the
previous step has to be finished to proceed with the next step. The
disadvantage of synchronous execution is the lower utilization of computational
resources since most of the GPUs won't perform any calculations while the Steps
2-4 are executed. The asynchronous case as shown in Figure~\ref{fig:asynch} is
designed to increase the utilization of computational resources by continually
updating the restarting configurations for the next MD step by continually
rerunning the analysis step. When one MD trajectory finishes, the MD worker
will retrieve the last generated restarting configuration for the next MD
trajectory and immediately start the next MD trajectory. This significantly
reduces the downtime for the MD workers. The disadvantage of the asynchronous
approach is that the analysis processes only a fraction of the MD trajectories
from the last iteration since these MD trajectories aren't finished when the
analysis starts. These trajectories will be fully analyzed only in the next
iteration. To reduce the unanalyzed length of the trajectories the analysis
step has to be continuously rerun, shown in Figure~\ref{fig:asynch}.  In the
work presented here, all the simulations are executed asynchronously. The
ability of asynchronous execution is a significant advantage over other
adaptive sampling packages.

\subsection{\label{sec:restart-strategies}Restart Strategies for Adaptive Sampling}

\begin{figure}[h!]
  \includegraphics[width=0.48\textwidth]{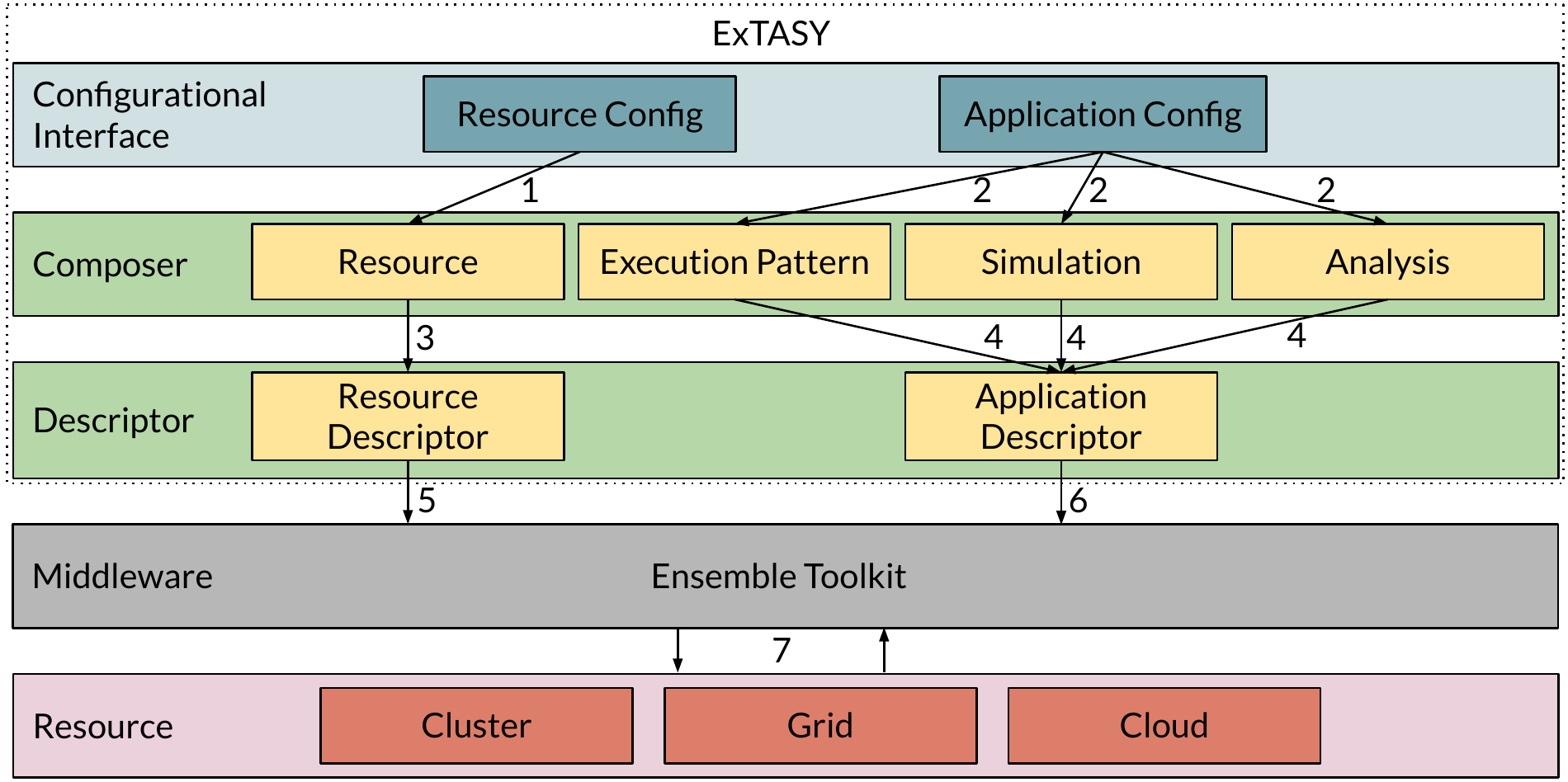}
  \caption{ExTASY-EnTK Integration: The diagram illustrates
  the seven execution steps of an adaptive sampling algorithm using ExTASY.  
  }\label{fig:extasy_arch}
\end{figure}

In the ExTASY framework, the different Restart Strategies in Step 4 in
Figure~\ref{fig:schema} are easily exchangeable. Here we use two
strategies: the first one is based on the count of the number of macrostates (hereby indicated as $cmacro$
strategy); the second one is based on the count of the number of microstate (hereby indicated as $cmicro$
strategy). The $cmacro$ strategy was shown to be more effective in reaching the
folded state of a protein from the unfolded state and the $cmicro$ strategy was shown to be more
effective in exploring the whole protein landscape \cite{Adstrategies2018}. 
These strategies do not assume any a priori knowledge of the system except
the structure of the starting configuration in the unfolded ensemble, but other adaptive sampling
strategies that use additional information about the protein can be used in
the ExTASY framework (e.g., \cite{FAST-Bowman-2015}).

\paragraph{Adaptive sampling strategy $cmicro$}
One simple restart strategy is starting new molecular dynamics trajectories in
the microstates which have the worse statistics, that is, that have been visited the least during prior iterations
\cite{weber2011characterization, Fabritiis-2014, AdaptivePELE-Lecina2017,
doerr2016htmd}. This statement can be quantified by using the counts in the count matrix of the MSM from Step 2,
that report on how many times all previous trajectories have visited each
microstate.  The probability that any given microstate is selected in Step 4 for
the batch of restart conformations is set as inversely proportional to its
associated count. The $cmicro$ strategy is effective in quickly exploring
new regions of the whole protein landscape and to better sample the protein
dynamics \cite{Adstrategies2018}.

\paragraph{\label{sec:macro} Adaptive sampling strategy $cmacro$} 
Another popular restart strategy for adaptive sampling is a macrostate-based
method indicated here as $cmacro$. The main advantage of this
method is the faster folding of proteins or crossing of transition barriers
\cite{Adstrategies2018}. This advantage is achieved by using eigenvectors of
the on-the-fly MSM from Step 2 to select more restart configurations in areas which are
kinetically disconnected or less explored. In this method, the microstates of
the on-the-fly MSM are clustered into macrostates, for
example with PCCA \cite{roblitz2013fuzzy}. Any microstate not connected in the
main MSM is treated as an additional macrostate. The number of macrostates can
be either fixed, as in this work, or determined based on the number of
slow processes emerging from the analysis. The macrostate count is determined by
measuring how many times any previous trajectory has visited each macrostate.
The restart conformations for the next iteration of adaptive sampling are then
chosen from each macrostate inversely proportional to the macrostate count.
Individual conformations within a macrostate are selected inversely
proportional to the microstate count within the macrostate.

\subsection{\label{sec:Tools}Tools and Software}
\begin{figure}[h!]
    \begin{lstlisting}
    from @radical.entk@ import @Task, Stage, Pipeline@
    p = Pipeline()

    sim_stage = Stage()   
    sim_task = Task()
    sim_task.executable = @<executable>@ #example openmm
    sim_task.arguments = <args> #example openmm args
    <add other task properties>
    sim_stage.add_tasks(sim_task)
    
    ana_stage = Stage()
    ana_task = Task()
    ana_task.executable = @<executable>@ #example pyemma
    ana_task.arguments = <args> #example pyemma args
    <add other task properties>
    
    ana_stage.add_tasks(ana_task)
    ana_stage.post_exec = {
        'condition': eval_sims(),
        'on_true':   add_sims(),
        'on_false':  terminate()
        }
    
    p.add_stages([sim_stage, ana_stage])
    \end{lstlisting}
    \caption{Pseudocode describing the adaptive sampling algorithm using the
    EnTK API}\label{extasy_snippet}
\end{figure}

ExTASY is a domain-specific workflow
system~\cite{turilli2019middleware,turilli2018building} for adaptive sampling
algorithms on HPC platforms. ExTASY exposes domain-specific parameters and
simulation configurations, but abstracts complexities of execution management,
resource acquisition, and management using \rct
(RCT)~\cite{turilli2019middleware}. RCTs are software systems designed and
implemented in accordance with the building blocks approach
\cite{turilli2018building}. Each system is independently designed with
well-defined entities, functionalities, states, events, and errors.
Specifically, ExTASY uses Ensemble Toolkit
(EnTK)~\cite{balasubramanian2018harnessing} and RADICAL-Pilot
(RP)~\cite{merzky2018using}. Ensemble
Toolkit~\cite{entk-icpp-2016,balasubramanian2018harnessing} provides the
ability to create and execute ensemble-based applications with complex
coordination and communication but without the need for explicit resource
management. EnTK uses RP~\cite{merzky2018using}, which provides resource
management and task execution capabilities. In this section, we describe the
ExTASY framework and how it leverages capabilities offered by EnTK and RP.

\subsubsection{ExTASY}

ExTASY exposes configuration files to interface with users and two components:
Composer and Descriptor.

The Composer validates the user input and creates the Resource, Execution
Pattern, Simulation, and Analysis sub-components. The Resource represents a
valid resource description; Execution Pattern describes the number of
iterations, number of simulation tasks per iteration and number of analysis
tasks per iteration; Simulation and Analysis describe the parameters to be
used for simulation and analysis tasks.

The Descriptor interfaces with Ensemble Toolkit, the execution middleware. It
consists of two sub-components: Resource Descriptor and Application
Descriptor. The former converts the resource description to a format as
accepted by the middleware. The latter uses the information from the Execution
Pattern, Simulation and Analysis sub-components to describe the complete
application to be executed.

Figure~\ref{fig:extasy_arch} presents the integration between ExTASY and the
execution middleware (EnTK). ExTASY translates the adaptive sampling
application into ordered executable tasks through a series of events: ExTASY
parses the configurational files to determine parameters to be used and
creates Resource description (event 1) and the simulation and analysis tasks
to be executed (event 2). ExTASY then uses EnTK's interface to describe the
resource and application (event 3 and 4) and initiate execution on the target
resource (event 5 and 6). EnTK executes all the simulations and analysis on
the resource (event 7).

ExTASY uses EnTK programming abstractions and EnTK's application programming
interface (API). ExTASY also uses EnTK's capabilities to support adaptive
execution~\cite{sncs20vivek} by modifying the execution plan
depending upon intermediate results e.g., add more simulations and analysis
tasks. Figure~\ref{extasy_snippet} provides pseudo-code on how ExTASY
implements an adaptive sampling algorithm using the EnTK API.

\subsubsection{Ensemble Toolkit}

EnTK simplifies the creation and execution of applications with complex
ensemble coordination and communication requirements. EnTK decouples the
description of ensemble coordination and communication from their execution by
separating three distinct concerns: (i) specification of task and resource
requirements; (ii) resource acquisition and management; and (iii) task
execution.

EnTK enables the encoding of ensemble applications by exposing an API with
four components: Application Manager, Pipeline, Stage and Task. Users specify
their application using pipelines, stages, and tasks. Users then pass this
specification and description of the target resource to the Application
Manager. Resource description includes properties like wall-time, number of
nodes and credentials for resource access.

The Task component is used to encapsulate an executable and its software
environment. The Stage component contains a set of tasks without mutual 
dependencies and that can therefore be executed concurrently. The Pipeline 
component is used to describe a sequence of stages. Description of ensemble
applications in terms of concurrency and sequentiality avoids the need to 
explicitly specify dependencies between tasks.

EnTK supports an explicit definition of pre and post conditions on the
execution of tasks, enabling fine-grained adaptivity~\cite{sncs20vivek}.
Adaptivity allows modifications to the number, type and order of tasks to be
executed during runtime, based on intermediate results. Specifically, EnTK
supports three types of adaptivity: (i) adaptivity in the number of tasks;
(ii) adaptivity in the order of tasks; and (iii) adaptivity in the properties
of a task.

EnTK provides a simple programming model, abstracts the complexities of
resource and execution management, and adds only a small and well-bounded
overhead on the execution O(1000) tasks~\cite{balasubramanian2018harnessing}.
EnTK uses a runtime system, such as RADICAL-Pilot, to acquire the resources
needed, manage task execution, as well as provide portability across
heterogeneous HPC resources.

\subsubsection{RADICAL-Pilot}

Two methods traditionally used to execute multiple HPC tasks are: (i) each
task is scheduled as an individual job; or (ii) use message-passing interface
(MPI) capabilities to execute multiple tasks as part of a single job. The
former method requires each task to be independently executed; the latter
method is suboptimal for heterogeneous or adaptive execution of tasks. The
pilot abstraction~\cite{turilli2018comprehensive} addresses some of these
limitations.  The pilot abstraction: (i) uses a placeholder job without any
tasks assigned to it, to acquire resources; and, (ii) decouples the
initial resource acquisition from task-to-resource assignment. Once the pilot
is scheduled, tasks are scheduled within its spatio-temporal resource
boundaries, which allows computational tasks to be executed directly without
being queued. The pilot abstraction thus supports the requirements of
high-throughput and task-level parallelism, while providing flexible execution
of tasks. RADICAL-Pilot~\cite{merzky2018using} is an implementation of the
pilot abstraction, engineered to support scalable and efficient launching of
heterogeneous tasks across different platforms.

\subsubsection{\label{sec:scaling} Scaling}

\begin{figure}[h]
  \centering
  \includegraphics[width=0.9\linewidth]{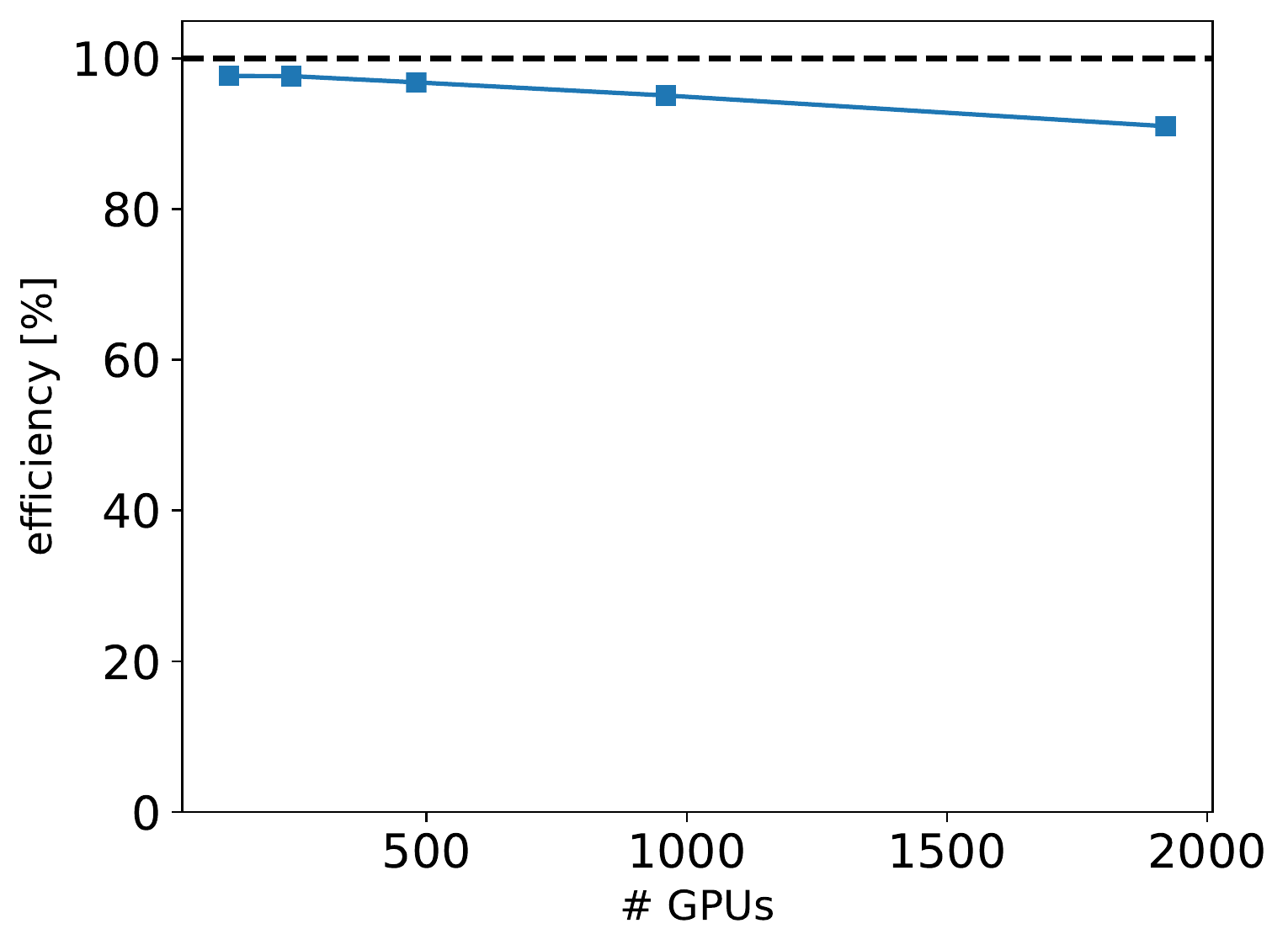}
  \caption{Scaling of ExTASY Efficiency on Summit. Efficiency was measured
  by running on all but one of the GPUs molecular dynamics tasks and on one GPU
  an analysis task, with 6 GPUs per node and a total length of the Summit jobs
  of 2 hours. The asynchronous, concurrent execution of the individual
  molecular dynamics and analysis tasks improves the efficiency of ExTASY
  compared to the previous version.\cite{Extasy2016}}
  \label{fig:scaling}
\end{figure}

ExTASY provides the seamless capability to
execute thousands of MD tasks on HPC, and provides capabilities to support
their interaction. ExTASY does not require modifications to the MD executables
and thus the performance of every MD executable remains unchanged. Adaptive
sampling on HPC platforms presents several performance
challenges~\cite{sncs20vivek} however, and a careful analysis of scaling
properties is critical. The use of ExTASY for complex workflow management,
task execution, and coordination introduces some overhead. The exact overhead
($\Omega$) depends on scale, and can be measured as a function of either the
number of tasks executed, or number of nodes employed. We use efficiency to
measure and quantify the scalability of ExTASY. Efficiency is defined as: $1 -
\Omega/2$ hours, where 2 hours is the total runtime of the scaling workload. The overhead numbers
reported are the aggregated overhead of individual RADICAL-Cybertools (EnTK
and RADICAL-Pilot) overhead, as well as the ``thin'' ExTASY layer. The primary
contribution to the overhead arises from the task coordination, placement and
execution -- functionality provided by RADICAL-Cybertools.


The efficiency is plotted in Figure~\ref{fig:scaling}. For every run,
all GPUs but one are set to run molecular dynamics tasks; one GPU is assigned
to run analysis tasks. Overheads increase modestly as the number of GPUs
(nodes) employed increases, which as shown in Figure~\ref{fig:scaling}, resulting
in an efficiency that is $> 90\%$ at $\approx$ 2000 GPUs on Summit.
This efficiency is: (i) agnostic of specific task executables (e.g., Gromacs,
AMBER, OpenMM etc.), (ii) not affected by the scalability of the individual
tasks or their runtime duration, and, (iii) dependent upon
RADICAL-Cybertools overheads, which are essentially invariant across different
HPC platforms\cite{turilli2019ac}.

\subsection{\label{sec:Reference} Reference Data}

\begin{table}[h!]
\centering
\caption{Adaptively sampled proteins in this study}
\label{tab:dataset-summary}
\resizebox{\columnwidth}{!}{
\begin{tabular}{|c|c|c|c|c|c|}
\hline
Protein & PDB ID & \# Residues & Folding Time ($\mu$s) \cite{lindorff2011} & Unfolding Time ($\mu$s)\\ 
\hline
Chignolin    & 5AWL                       & 10                 & 0.6                &2.2            \\
Villin       & 2F4K                       & 35                 & 2.8                &0.9            \\
BBA          & 1FME                       & 28                 & 18                 &5              \\
A3D          & 2A3D                       & 73                 & 27                 &31              \\
\hline            
\end{tabular}
}
\end{table}
To show the speed up and accuracy of the adaptive sampling method we projected
the results of the 4 small proteins on to preexisting long MD simulations, obtained
on the Anton supercomputer \cite{lindorff2011}. These proteins and the
reference data were investigated before \cite{reanalyze1, reanalyze2},
allowing us to demonstrate here the usefulness and reliability of the ExTASY
workflow. The 4 proteins are summarized in Table \ref{tab:dataset-summary},
their size is from 10 to 73 residues and have a short folding time below 40 $\mu$s.
These proteins were chosen due to their folding time which is long enough to
show the advantages of adaptive sampling with ExTASY framework, but still
reachable with our computational resources. Only the C-alpha coordinates are
used when comparing the reference data trajectories with the results from the
ExTASY framework in this work. Since parallelization strongly affects the
time to fold, we compare the results of adaptive sampling with the results of
plain MD with the same number of parallel simulations, starting from the same
starting configuration. Both adaptive sampling and plain MD were executed with
the ExTASY platform.

\subsection{\label{sec:MD} Molecular dynamics simulation}

\begin{figure}[h!]
   \begin{subfigure}[b]{0.85\linewidth}
   {\scalebarimg{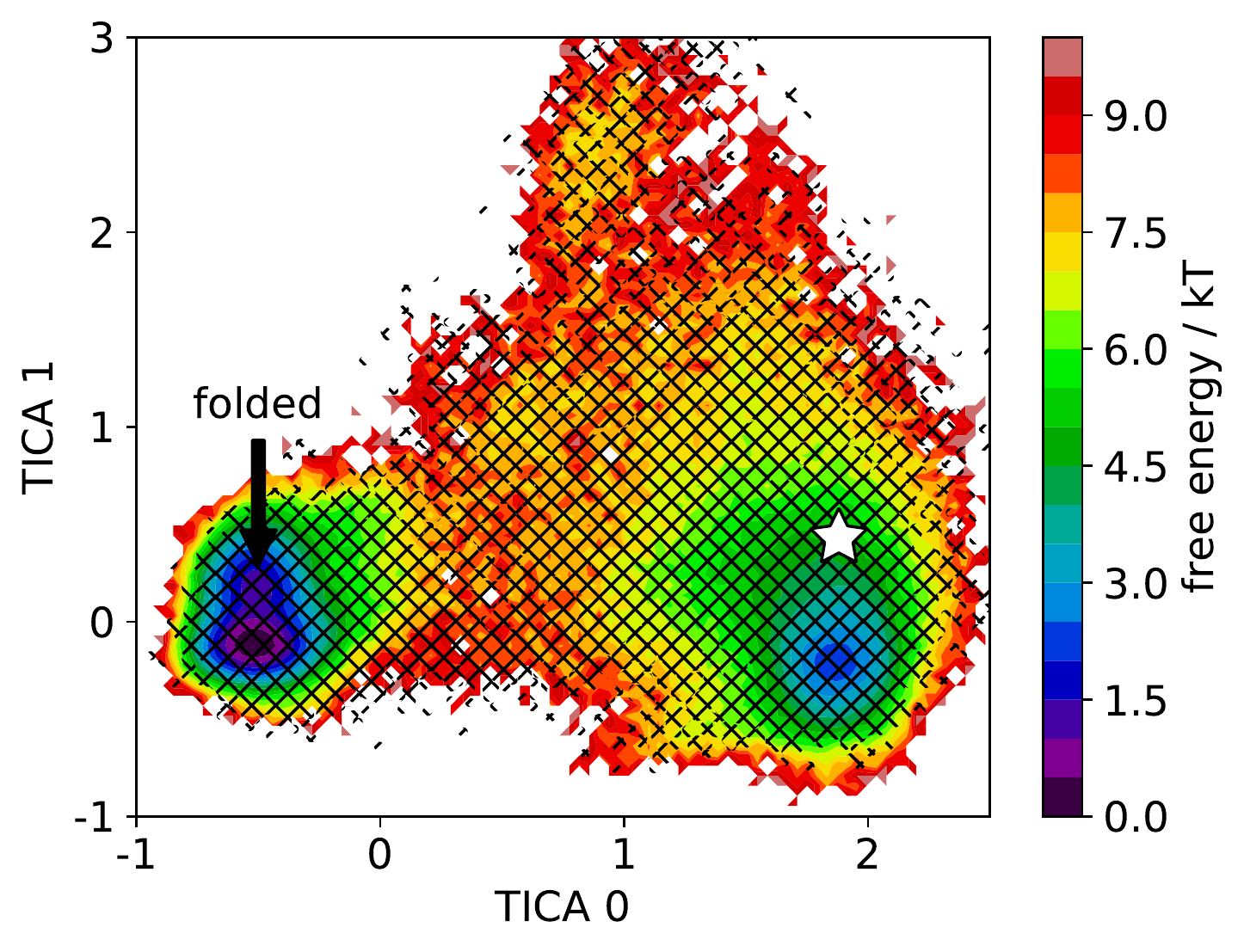}{0.98}{A)}{0mm}{-2mm}}
   \end{subfigure}%
   
   \begin{subfigure}[b]{0.85\linewidth}
   {\scalebarimg{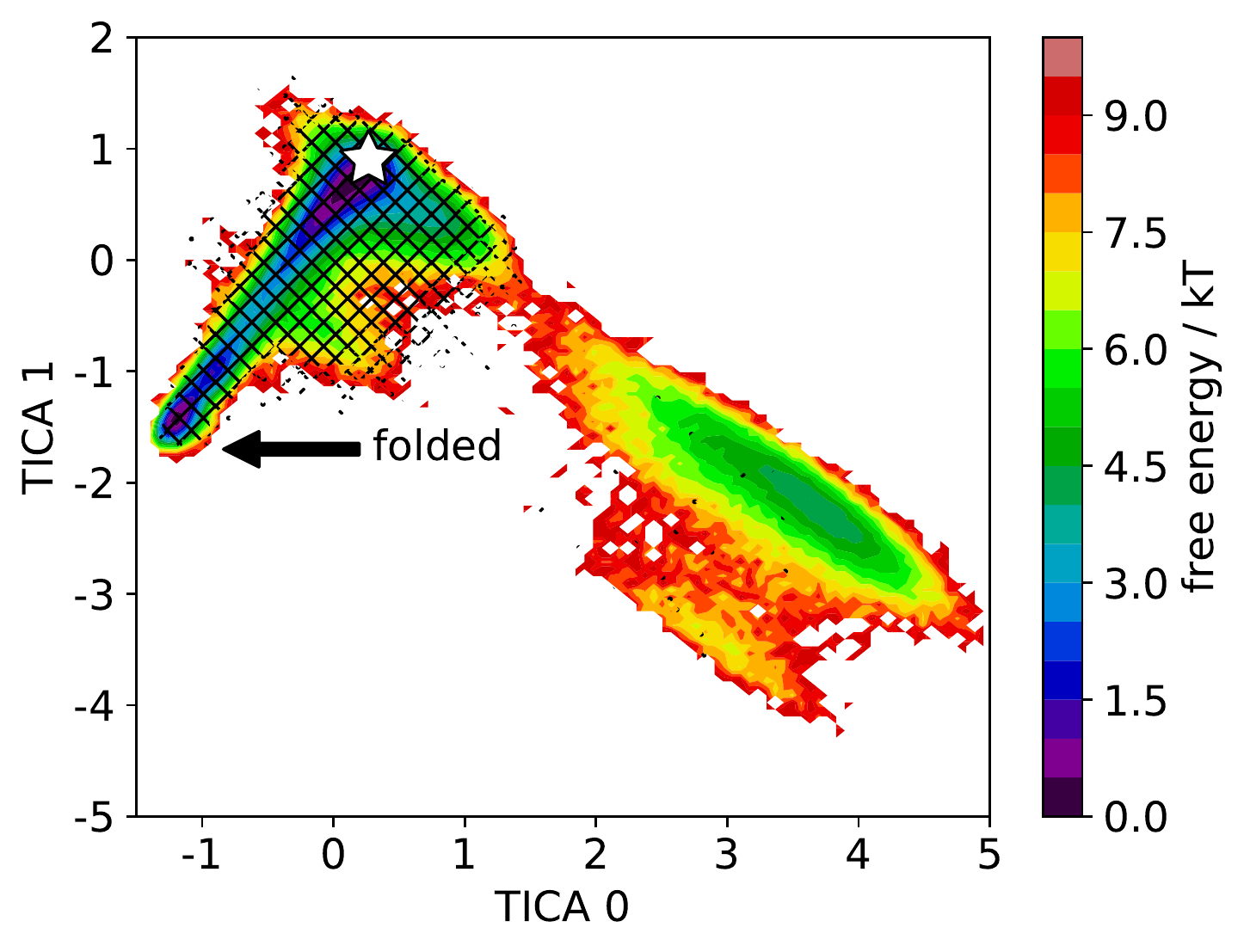}{0.98}{B)}{0mm}{-2mm}}
   \end{subfigure}%

  \caption{Exploration of the protein energy landscape in TICA coordinates. The
 color background shows the explored Free Energy landscape by the reference
 dataset. The black diagonal lines on top show the explored conformations by the adaptive
 sampling in this work. The almost perfect overlap shows that the whole
 conformational landscape of all 4 proteins was fully explored. The labels
 show the location of the folded states. The stars show the initial unfolded
 states. Individual proteins: A) Chignolin B) Villin }
\end{figure}

\begin{figure}[h!]\ContinuedFloat

   \begin{subfigure}[b]{0.85\linewidth}
   {\scalebarimg{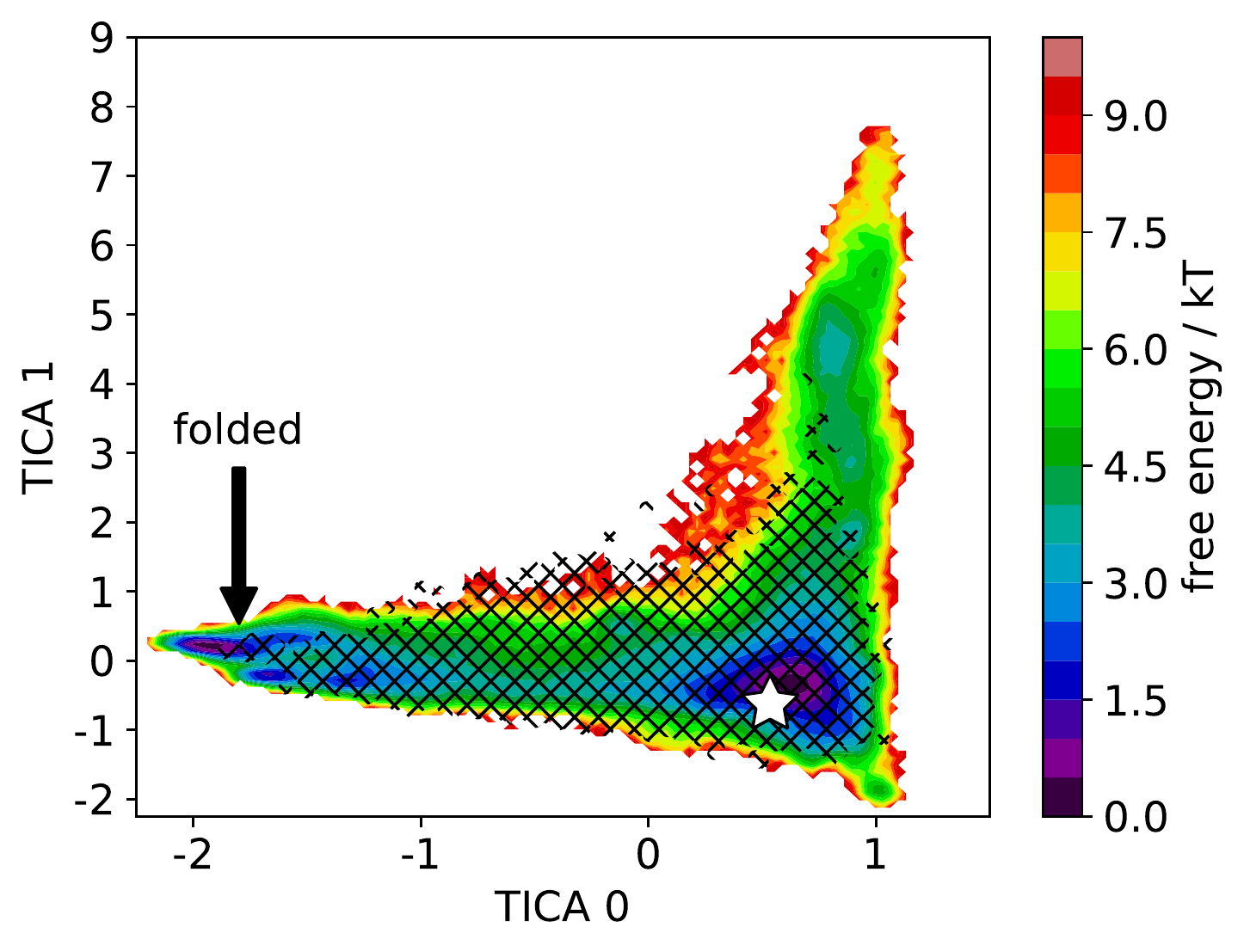}{0.98}{C)}{0mm}{-2mm}}
   \end{subfigure}%
  
   \begin{subfigure}[b]{0.85\linewidth}
   {\scalebarimg{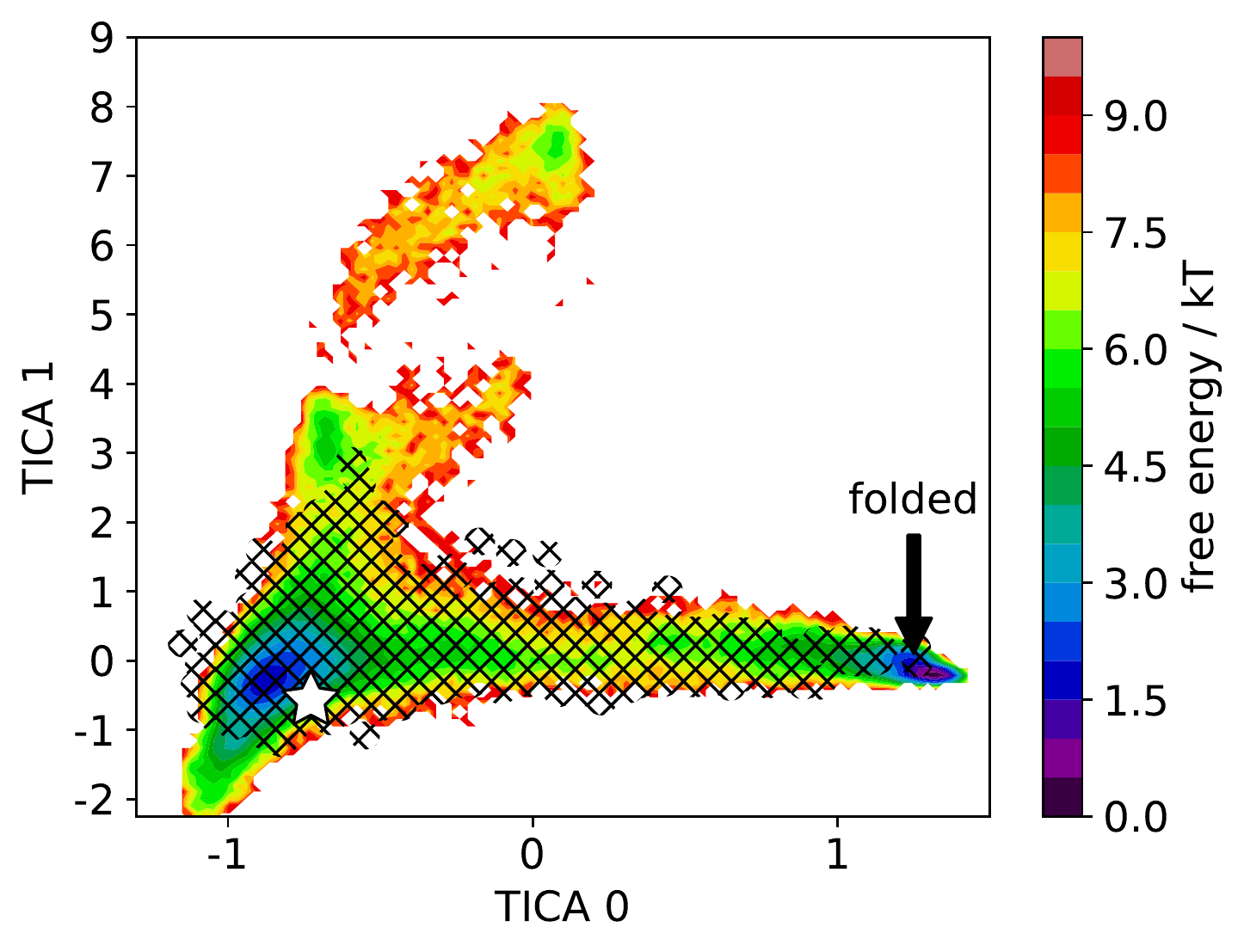}{0.98}{D)}{0mm}{-2mm}}
   \end{subfigure}%

  \caption{(cont.) Individual proteins: C) BBA D) A3D} 
  \label{fig:overlap}
\end{figure}

The MD simulations in this work were performed with OpenMM 7.5 \cite{openMM}
using CUDA 9.1 on the Summit supercomputer. To reproduce the same setup
as in \cite{lindorff2011} we used the CHARMM22* force field \cite{Charmm22star}
and the modified TIP3P water. The stepsize used was 5fs or 2fs in case of
protein A3D, and the trajectories were strided to to reduce the data
volume. Differently
from \cite{lindorff2011}, we used the Particle Mesh Ewald method for
long-range electrostatics due to OpenMM settings. For each protein, the start configuration
is one frame in the reference dataset selected randomly from the 20\%
of frames with the highest Root Mean Square Deviation (RMSD) from the
protein crystal structure. A short energy equilibration (1-2ns) was then
performed in the NPT ensemble to create initial coordinates for the workflows.
No further a priori information was given to the ExTASY framework except the
unfolded start conformations.

In each iteration, 50 OpenMM trajectories were simulated on 50 GPUs on
Summit with one GPU per trajectory. The length of each trajectory was 50ns
for Chignolin and Villin, 10ns for BBA and 40 ns for A3D. ExTASY scales up to
1000s of GPUs, so it can be used to simulate even larger proteins or a larger
number of parallel walkers.  Steps 2 and 4 were performed on one GPU on
Summit utilizing the same job as the MD simulations.  After all the
simulations are finished, the folding times, speed up and the accuracy of
protein dynamics was determined by comparing with the Anton MD simulations
starting from the selected start
conformation. 
We illustrate the abilities of adaptive sampling and the ExTASY framework
by comparing the results with what obtained with plain MD (i.e., without adaptive sampling) for the 4 proteins.
The variation of the time to fold for both adaptive sampling and plain
molecular dynamics can reach 50\% \cite{Adstrategies2018}, caused by
stochasticity.

\begin{figure}[h!]
   \begin{subfigure}[b]{0.8\linewidth}
   {\scalebarimg{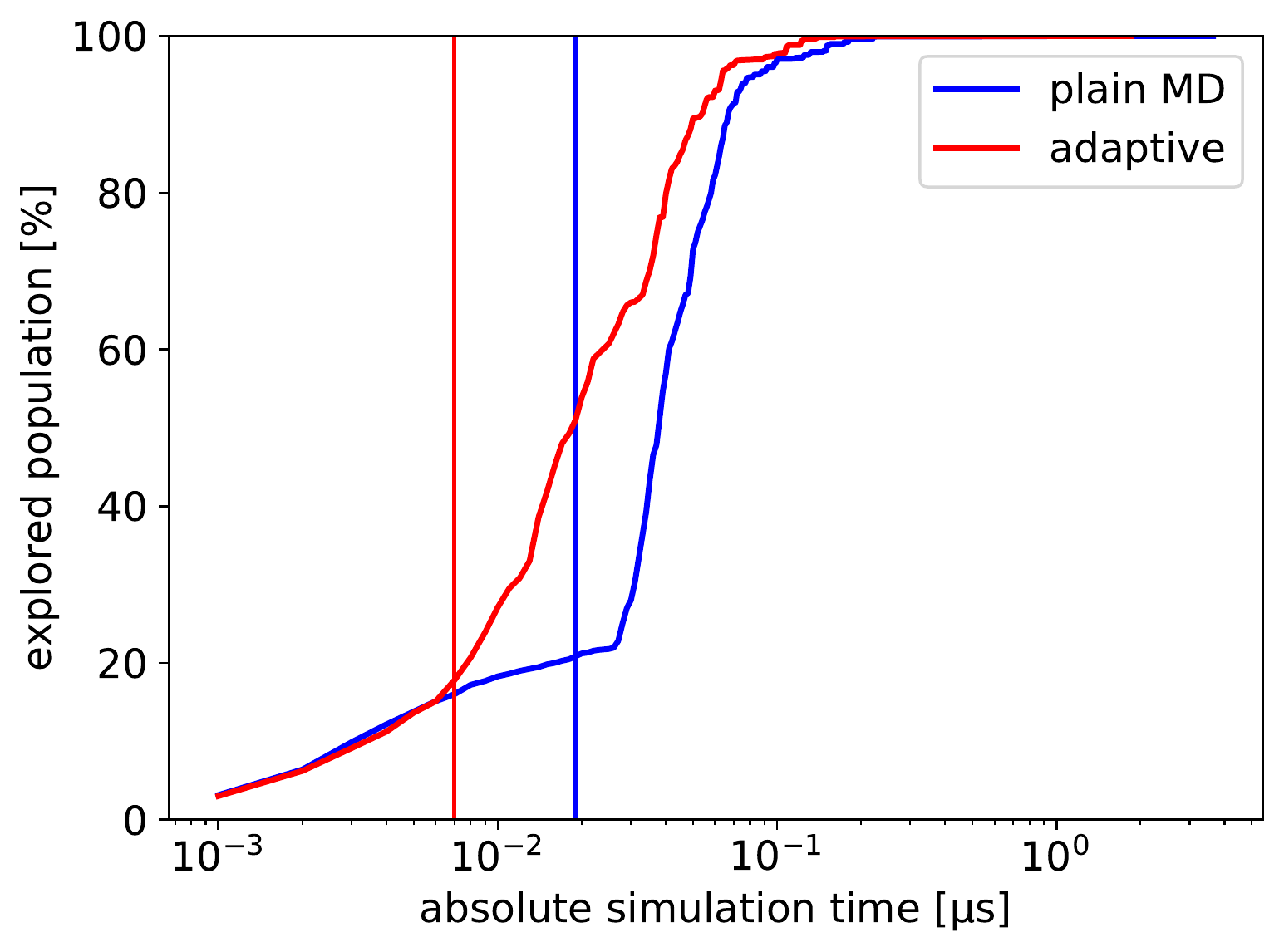}{0.95}{A)}{0mm}{-5mm}}
   \end{subfigure}%
   
   \begin{subfigure}[b]{0.8\linewidth}
   {\scalebarimg{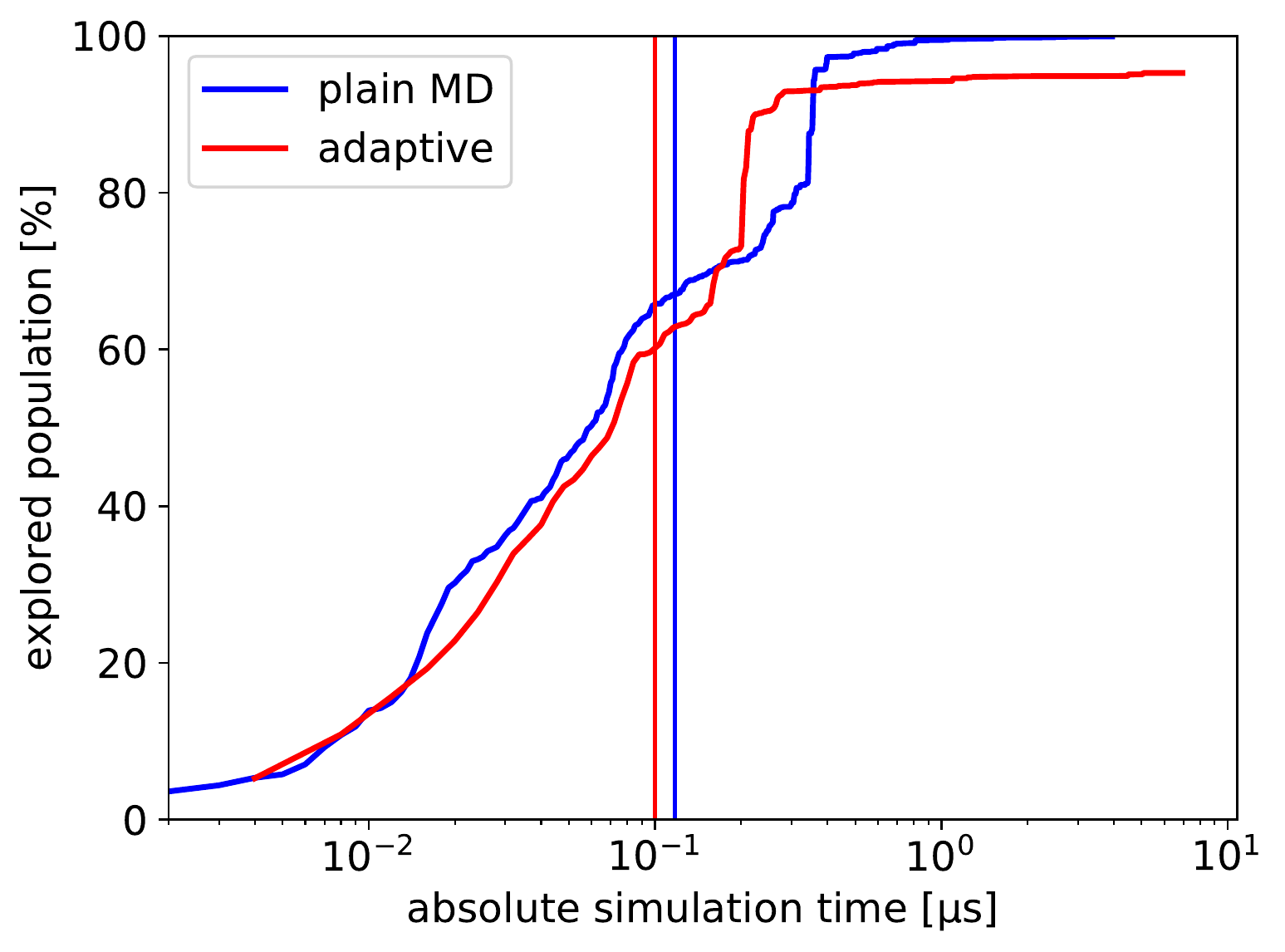}{0.95}{B)}{0mm}{-5mm}}
   \end{subfigure}%

  \caption{The population of explored states evolving with absolute simulation time.  Around one
  order of magnitude shorter time to solution can be reached with adaptive
  sampling compared to plain MD simulations. For Chignolin and A3D the
  $cmacro$ adaptive sampling strategy was used, for Villin and BBA the $cmicro$
  adaptive sampling strategy was used. Both adaptive sampling strategies are
  described in Section~\ref{sec:restart-strategies}. The vertical lines
  indicate folding events. Individual proteins: A) Chignolin B) Villin }
\end{figure}

\begin{figure}[h!]\ContinuedFloat
   \begin{subfigure}[b]{0.8\linewidth}
   {\scalebarimg{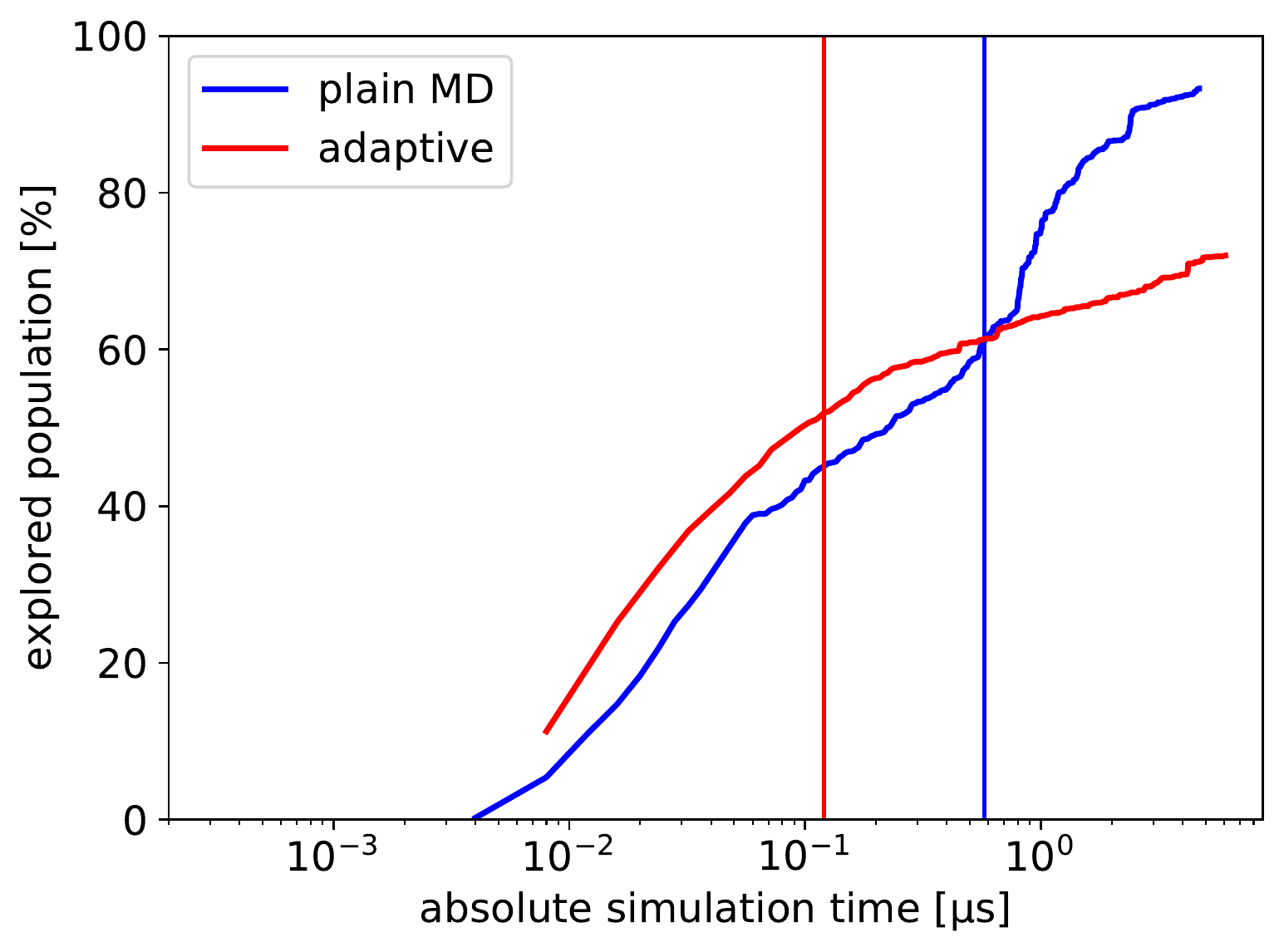}{0.95}{C)}{0mm}{-5mm}}
    \end{subfigure}%

   \begin{subfigure}[b]{0.8\linewidth}
   {\scalebarimg{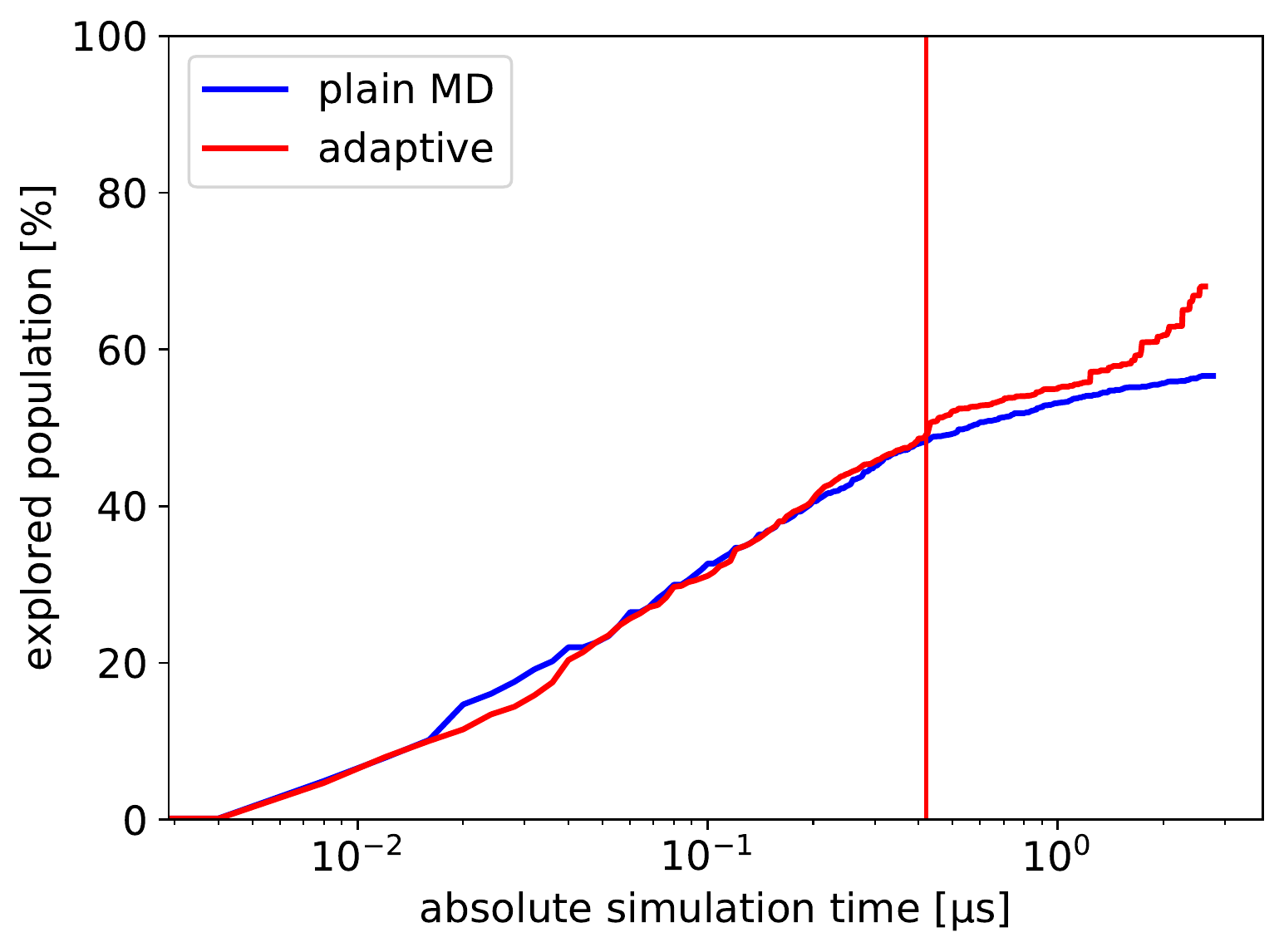}{0.95}{D)}{0mm}{-5mm}}
    \end{subfigure}%
  \caption{
  (cont.) Proteins C) BBA D) A3D  The plain molecular dynamics
  simulation of A3D has not folded despite simulating about 7 times longer than
  necessary for folding with adaptive sampling. Additional computational
  resources would allow to fold protein A3D with plain molecular dynamics too.}
  \label{fig:Pop_explored}
\end{figure}
\section{\label{sec:results}Results and Discussion}

To analyze the efficiency of adaptive sampling, we considered several
measures. To show the completeness of the exploration, we measured the fraction
of the explored population and considered the overlap of the explored areas with the
reference dataset. This also allows us to estimate the speed up time to solution
compared to the reference method. To analyze the accuracy of the simulated
protein dynamics we compare the relative entropy of the MSM transition
matrices, and the Mean First Passage Time (MFPT) to the folded state, as
detailed below.

\subsection{\label{sec:time-fold}Comparison of Exploration}
\begin{figure}[!htb]
   
   \begin{subfigure}[b]{0.85\linewidth}
   {\scalebarimg{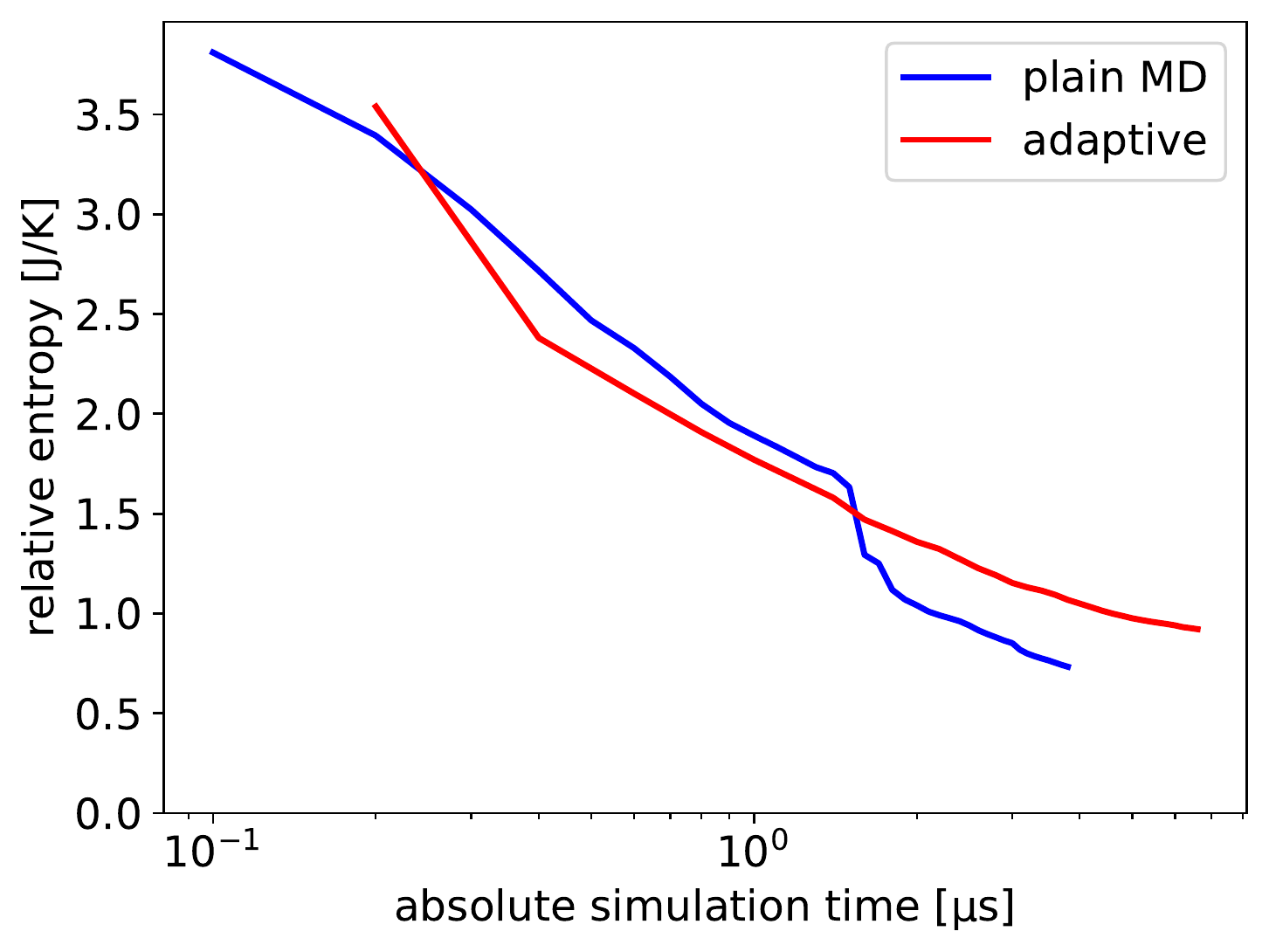}{1}{}{0mm}{-3mm}}
   \end{subfigure}%
   
  \caption{
  Relative entropy between the MSM transition matrices generated during the
  ExTASY exploration and from the plain MD comparison run. Results
  for protein Villin. The relative entropy decreases with increasing number of adaptive
  sampling iterations. }
  \label{fig:rel_ent}
\end{figure}

The whole explored energy landscape of the protein cannot be visualized due to
the high dimensionality of the raw trajectories, but the explored landscape in
the reduced TICA coordinates is shown in
Figure~\ref{fig:overlap}. The colored background shows the explored free energy
landscape of the reference dataset and the shaded foreground shows the region
of this landscape explored by the adaptive sampling. The whole energy
landscapes of the proteins Chignolin, Villin, BBA and A3D were explored by
ExTASY. The small differences in overlap could be caused by the differences in
the long-range electrostatics setup and the stochastic nature of the
exploration. The case of protein A3D shows that for larger protein the
computational resources to fully explore the energy landscape rise
significantly. Our computational resources allowed us to fold A3D with
adaptive sampling, but not with plain molecular dynamics. The plain molecular
dynamics simulation didn't fold even when simulating about 7 times longer than
the time to fold with adaptive sampling. To show that adaptive sampling works
with different adaptive sampling strategies, we folded Chignolin and A3D with
the $cmacro$ adaptive sampling strategy while the $cmicro$ adaptive sampling
strategy was used to fold Villin and BBA.

To show how effectively the whole protein landscape is explored, we
use the fraction of the total population explored as a function of time.
Here we select all the states which are explored at a certain time and
compare with all possible states (as obtained from the reference simulations).
To represent the different importance of different states we weight the
explored states with their stationary weight. The population of each microstate
is calculated as the stationary weight of that microstate from the MSM analysis of
the reference dataset. Figure~\ref{fig:Pop_explored} reports the comparison of
the explored populations as a function of time for the reference dataset and
the ExTASY results. For all 4 proteins, adaptive sampling explores the
protein energy landscape slightly faster and folds significantly faster. The
folding speed up of adaptive sampling compared to plain MD is 170\% for
Chignolin, 20\% for Villin, 380\% for BBA, and more than 690\% for A3D. The
larger speed up for A3D is in line with the prediction that for larger and more
complex proteins, adaptive sampling achieves a larger speed
up\cite{Adstrategies2018}. The exact speed up for the protein A3D could not be
determined as we didn't observe any folding event with plain MD.

The x-axis in Figures~\ref{fig:Pop_explored}-\ref{fig:mfpt}
reports the absolute simulation time to show the improvement of time to solution with ExTASY.
The absolute simulation time is the length of one trajectory in Step 1 times the number of iterations.
When all the trajectories in Step 1 are run in parallel, the absolute simulation time
shows the time to solution independent of the used hardware. The effects of parallelization
on the time to solution for adaptive sampling were explored in previous work\cite{Adstrategies2018},
generally parallelization decreases the time to solution.

\begin{figure}[!h]
   \begin{subfigure}[b]{0.87\linewidth}
   {\scalebarimg{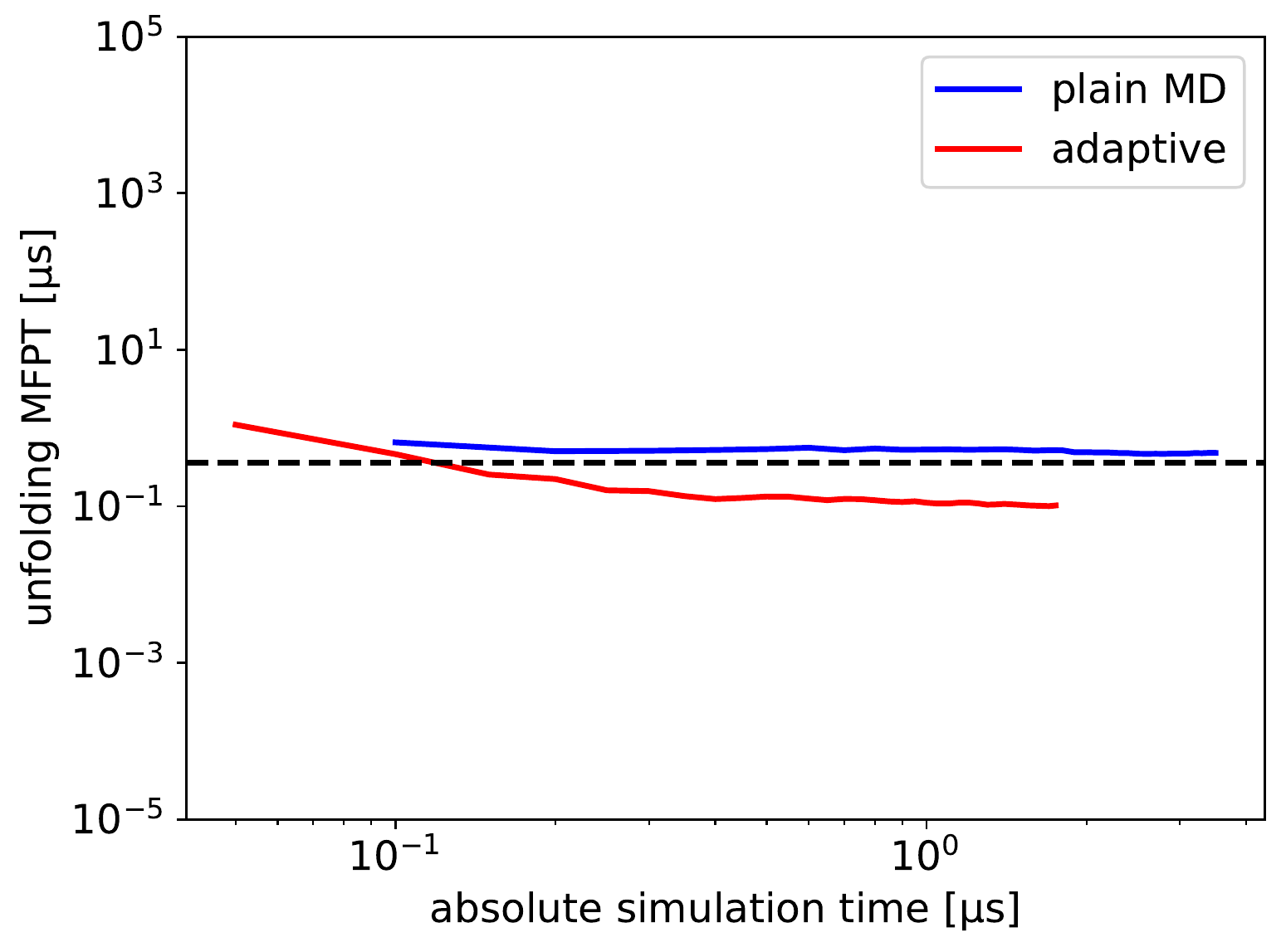}{1}{A)}{0mm}{-2mm}}
   \end{subfigure}%
   
   \begin{subfigure}[b]{0.87\linewidth}
   {\scalebarimg{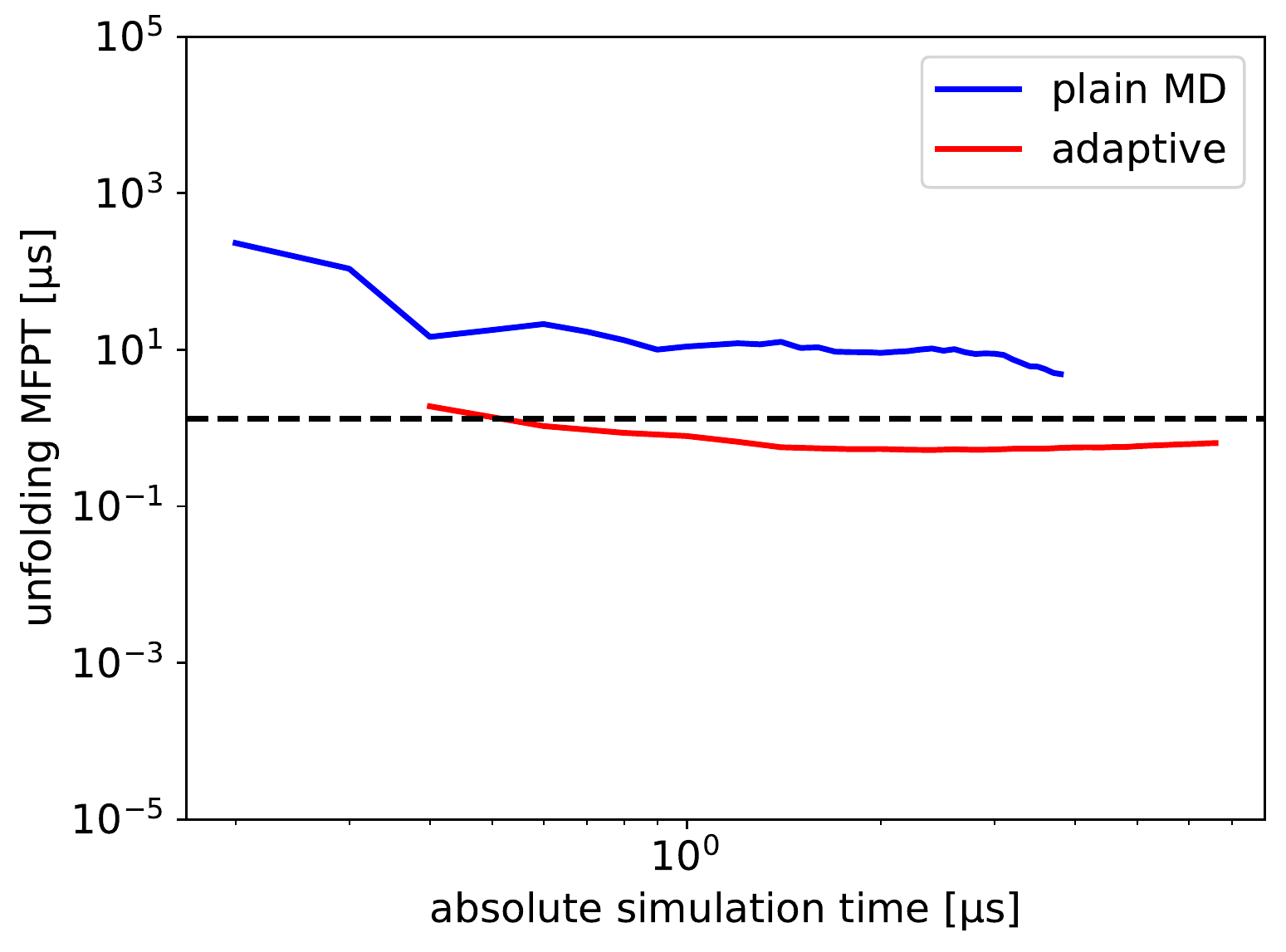}{1}{B)}{0mm}{-2mm}}
   \end{subfigure}%
   
  \caption{
  MFPT from folded to unfolded states evolving as
  more data is available after more adaptive sampling iterations. The red line
  corresponds to the adaptive sampling results, the blue line corresponds to plain MD with the same parallelization
  as adaptive sampling. The black dashed line shows the reference values from the
  Anton simulation trajectories\cite{lindorff2011}.   
  Individual proteins: A) Chignolin B) Villin.} 
\end{figure}

\begin{figure}[!h]\ContinuedFloat

   \begin{subfigure}[b]{0.87\linewidth}
   {\scalebarimg{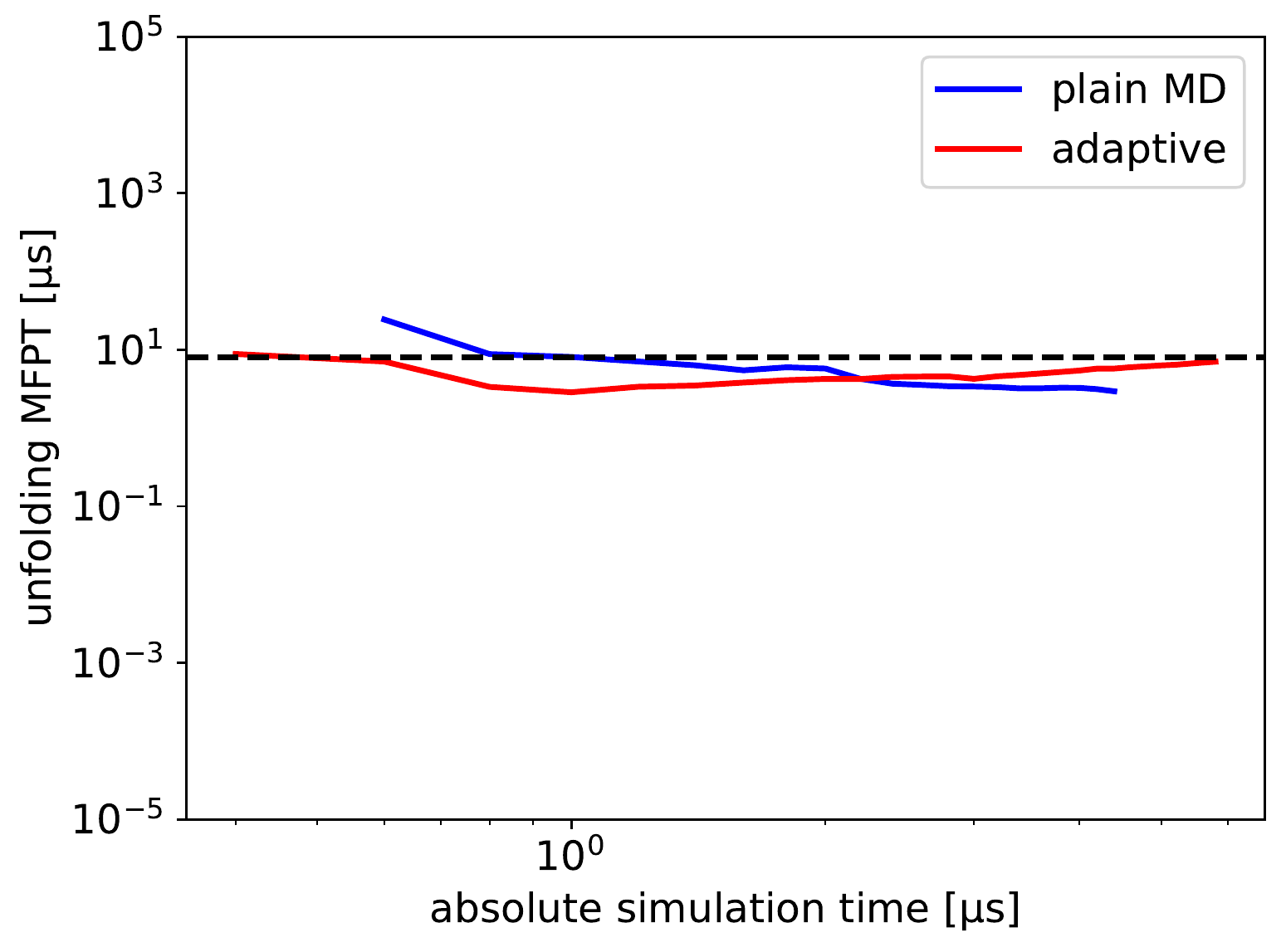}{1}{C)}{0mm}{-2mm}}
    \end{subfigure}%

    \begin{subfigure}[b]{0.87\linewidth}
   {\scalebarimg{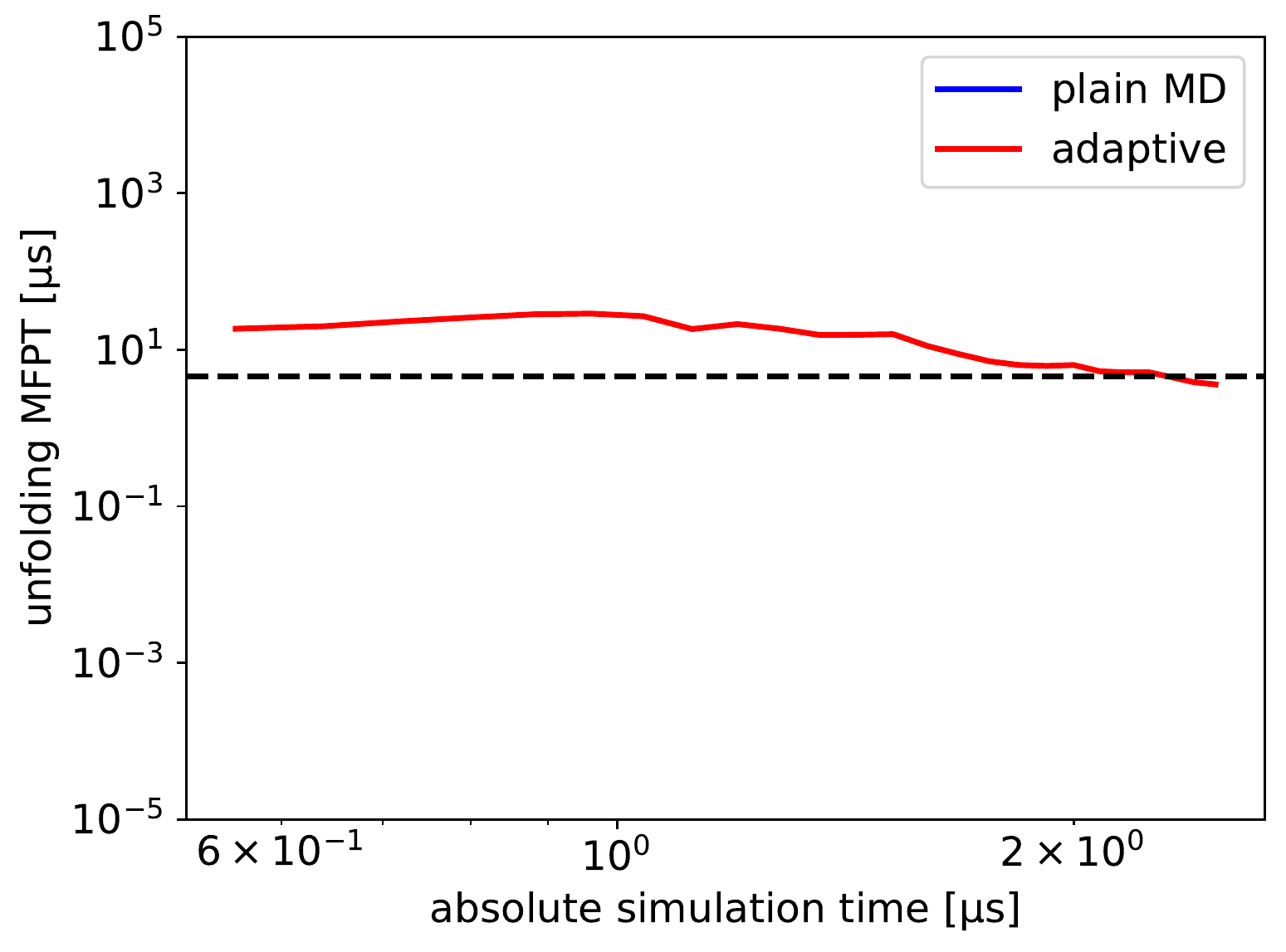}{1}{D)}{0mm}{-2mm}}
    \end{subfigure}%

  \caption{
  (cont.) Proteins: C) BBA D) A3D.
  The MFPT for A3D is available only for adaptive sampling since the plain
  molecular dynamics simulation did not produce any folding event, even while
  simulating for about 7 times longer with adaptive sampling.}   
  \label{fig:mfpt}
\end{figure}

\subsection{\label{sec:kinetics}Comparison of Protein Dynamics}

To track the convergence of protein dynamics in the adaptive sampling workflow,
one can use the relative entropy \cite{bowman2010enhanced} between the
MSM transition matrix of the reference data, $P_{ij}$, and the MSM transition matrix
of the analyzed data, $Q_{ij}$.
A relative entropy can be calculated between each microstate in the analyzed
and the reference transition probabilities from this state. By averaging the
relative entropy for each state weighted by the stationary probability over all
microstates we obtain the relative entropy between the two transition matrices.
The relative entropy
$D(P||Q)$ is then given by
\begin{equation}
D(P||Q)=\sum_{i,j}^{N}s_{i}P_{ij}\ln\frac{P_{ij}}{Q_{ij}}. 
\end{equation}
where $s_{i}$ is the equilibrium probability of state $i$. The transition
matrices $P_{ij}$ and $Q_{ij}$ are obtained by using exactly the same dimension reduction
and same clustering.
As zero counts in the transition matrices can cause divergence of the relative
entropy, a pseudo-count of $1/N$ (where $N$ is the length of the simulation) is
added to each element of the count matrices
before normalizing the rows to get the transition matrices
\cite{bowman2010enhanced}. The relative entropy for a certain simulation time is
obtained from all the trajectories up to the specified simulation time. By definition, the relative entropy of the full reference
trajectory is zero.  Figure~\ref{fig:rel_ent} shows how the relative
entropy decreases with increasing simulation time for both the adaptive
sampling and plain MD simulations. The adaptive sampling strategy decreases the
relative entropy faster at the beginning, later in the simulation plain MD
decreases the relative entropy faster. While the sample size is small, this
confirms that the chosen sampling strategy is effective at
exploring the protein landscape, but not optimized in the later steps of
converging protein kinetics \cite{Adstrategies2018}. Different adaptive
sampling strategies which are optimized for converging the kinetics could
improve the behavior of adaptive sampling towards the end of the simulation.

An additional measure to compare the convergence of the kinetic behavior
of the proteins is the Mean First Passage Time (MFPT). MFPT measures the mean time
to reach for the first time another state from one state. Here the two states
are the folded and unfolded states, both defined as an ensemble of MSM microstates
based on the position in the TICA coordinates' space. 
Figure~\ref{fig:mfpt} shows how the MFPT from
the folded to the unfolded state converges as a function of simulation time,
for both the adaptive sampling and the plain MD simulations. The reference MFPT
obtained from the Anton trajectories shows that the results of both plain MD and
adaptive sampling converge to the same order of magnitude of the reference value.
Adaptive sampling shows larger errors in the case of Chignolin, but a smaller
error in the case of Villin and BBA compared to plain MD. The small sample size
prevents us to conclude if plain MD or adaptive sampling converges faster for Chignolin, Villin, and BBA proteins. 
For protein A3D the convergence of MFPT could not be compared since the plain
MD simulation did not produce folding events, while the adaptive sampling
simulation converged to a MFPT value close to the reference value.
The adaptive sampling strategies used in this work are not optimized for
convergence of kinetics. The convergence of MFPT shows that, not surprisingly,
the amount of sampling required to reach accurate kinetic values is longer than the sampling
required to fold a protein. Kinetic values from both plain MD and adaptive
sampling have relatively large uncertainties. The sizes of the MFPT errors are
similar to what was obtained with the HTMD framework \cite{doerr2016htmd}.

\section{\label{sec:conclusion}Conclusion}

We have shown that the ExTASY framework \cite{Extasy2016} can effectively
perform adaptive sampling, as exemplified by simulations on 4
proteins using deep learning in the analysis step. The free energy landscape of the 4 proteins was
fully sampled. In comparison to a plain molecular dynamics simulation with the
same parallelization a statistically significant speed up in the
range of 20\% to 690\% could be observed. For the largest of the protein studied, A3D, folding could
be achieved with adaptive sampling but not with plain MD. The obtained speed ups are
in line with predictions \cite{Adstrategies2018} corresponding to the size of
the proteins studied.
The MFPT times converged for both adaptive sampling and plain MD to values
similar to the reference values. For protein BBA the adaptive sampling
converges the MFPT significantly faster than plain MD, but proteins Chignolin
and Villin show a slower convergence. 
The sample size in this paper was limited by computational
resources. Additional adaptive sampling strategies optimized in recovering the
kinetics would improve the results of adaptive sampling for MFPT convergence.
The relative entropy between the transition matrix of the MSM computed during
the adaptive
sampling and the MSM of the plain MD decreases steadily with the
simulation time, the adaptive sampling decreases here faster than plain MD at
the beginning of the simulation. The differences in speed of convergence are
caused by the choice of adaptive sampling strategy, which are validated for the
exploration of protein energy landscapes, but not optimized to reach accurate
protein kinetics.
The modularity of the ExTASY framework reduces the time spent by domain
experts in executing
adaptive sampling in a scalable fashion on diverse platforms. Scalability up to
2000 GPUs and 2000 simultaneous protein replicas was demonstrated on the Summit supercomputer.
This scalability doesn't come at the cost of inflexibility, due to the design
and implementation of the ExTASY framework \cite{Extasy2016}, ExTASY can be
easily modified for different proteins or MD
simulation software. ExTASY can also be easily extended to different adaptive
sampling strategies and platforms. The ExTASY framework is available
open-source at
\href{https://github.com/ClementiGroup/ExTASY}{https://github.com/ClementiGroup/ExTASY}.
The flexibility of ExTASY allowed this framework to be the first
open-source adaptive sampling platform which supports an analysis step based on
deep learning or asynchronous execution.

\section{Supplementary materials}
See Supplementary material for the ExTASY workflow parameters and more results
for the 4 proteins.

\section{Acknowledgments}
We thank D.E. Shaw Research for the reference dataset of Molecular Dynamics
trajectories. We thank Matteo Turilli, Andre Merzky and other members of the
RADICAL Team (http://radical.rutgers.edu) for their support with performance
analysis and RADICAL-Cybertools on Summit. We also thank John Ossyra (Oak
Ridge) for useful discussions. This work is supported in part by the National
Science Foundation (CHE-1265929, CHE-1740990, CHE-1900374, and PHY-1427654 to
C.C.), the Welch Foundation (C-1570 to C.C.). Supercomputing time was provided
by Blue Waters (supported by the National Science Foundation, awards
OCI-0725070 and ACI-1240993) sustained-petascale computing project via NSF
1713749. Additional Supercomputing time was provided on Summit, which is
a DOE Office of Science User Facility supported under Contract
DE-AC05-00OR22725.


\providecommand{\noopsort}[1]{}\providecommand{\singleletter}[1]{#1}%
\providecommand{\latin}[1]{#1}
\makeatletter
\providecommand{\doi}
  {\begingroup\let\do\@makeother\dospecials
  \catcode`\{=1 \catcode`\}=2 \doi@aux}
\providecommand{\doi@aux}[1]{\endgroup\texttt{#1}}
\makeatother
\providecommand*\mcitethebibliography{\thebibliography}
\csname @ifundefined\endcsname{endmcitethebibliography}
  {\let\endmcitethebibliography\endthebibliography}{}

\end{document}